\documentclass[12pt]{article}

\usepackage[margin=1in]{geometry}  	
\usepackage{amsfonts}
\usepackage{amsmath} 
\usepackage{booktabs}
\usepackage{multirow}
\usepackage[doublespacing]{setspace} 

\usepackage{float}
\usepackage{caption}
\usepackage{graphicx}
\usepackage{adjustbox}
\usepackage{lscape}

\usepackage{adjustbox}

\usepackage{xcolor}

\usepackage{relsize}

\usepackage{bbm}
\usepackage{subfig}

\definecolor{forestgreen}{RGB}{34,139,34}
\usepackage{hyperref}
\hypersetup{
    colorlinks=true,
    linkcolor=magenta,
    filecolor=magenta, 
    citecolor=forestgreen,      
    urlcolor=blue
}

\makeatletter
\newcommand{\vast}{\bBigg@{3}}
\newcommand{\Vast}{\bBigg@{4}}
\makeatother

\makeatletter
\newcommand*{\indep}{%
  \mathbin{%
    \mathpalette{\@indep}{}%
  }%
}
\newcommand*{\nindep}{%
  \mathbin{
    \mathpalette{\@indep}{\not}
  }%
}
\newcommand*{\@indep}[2]{%
  \sbox0{$#1\perp\m@th$}
  \sbox2{$#1=$}
  \sbox4{$#1\vcenter{}$}
  \rlap{\copy0}
  \dimen@=\dimexpr\ht2-\ht4-.2pt\relax
  \kern\dimen@
  {#2}%
  \kern\dimen@
  \copy0 
} 
\makeatother

\DeclareMathOperator{\E}{\mbox{E}}

\usepackage[utf8]{inputenc}
\DeclareUnicodeCharacter{200E}{}

\usepackage{authblk}

\usepackage{tikz}
\usetikzlibrary{positioning,shapes.geometric}

\usepackage{setspace}
\usepackage{cite}

\usepackage[stable]{footmisc}

\usepackage{rotating}

\makeatletter
\def\@seccntformat#1{\@ifundefined{#1@cntformat}%
   {\csname the#1\endcsname\quad}  
   {\csname #1@cntformat\endcsname}
}
\let\oldappendix\appendix 
\renewcommand\appendix{%
    \oldappendix
    \newcommand{\section@cntformat}{\appendixname~\thesection\quad}
}
\makeatother

\usepackage[absolute,showboxes]{textpos}

\TPMargin{5pt}

\newcommand{\copyrightstatement}{
    \begin{textblock}{0.84}(0.08,0.93)    
         \noindent
         \footnotesize
         This DRAFT manuscript presents WORK IN PROGRESS. Please send comments to \href{mailto:idahabreh@hsph.harvard.edu}{idahabreh@hsph.harvard.edu}.
    \end{textblock}
}

\usepackage{datetime}
\def\paperversionmajor{25}
\def\paperversionminor{0}

\begin{document}

\copyrightstatement

\title{Global sensitivity analysis for studies extending inferences from a randomized trial to a target population \vspace{0.25in}}

\author[1-3]{Issa J. Dahabreh\thanks{Address for correspondence: Dr. Issa J. Dahabreh; CAUSALab; Harvard T.H. Chan School of Public Health, Boston, MA 02115; email: \href{mailto:idahabreh@hsph.harvard.edu}{idahabreh@hsph.harvard.edu}; phone: +1 (617) 495‑1000.}}
\author[1-3]{James M. Robins}
\author[2,3]{Sebastien J-P.A. Haneuse}
\author[1,2]{Sarah E. Robertson}
\author[4]{Jon A. Steingrimsson}
\author[1-3]{Miguel A. Hern\'an}

\affil[1]{CAUSALab, Harvard T.H. Chan School of Public Health, Boston, MA}
\affil[2]{Department of Epidemiology, Harvard T.H. Chan School of Public Health, Boston, MA}
\affil[3]{Department of Biostatistics, Harvard T.H. Chan School of Public Health, Boston, MA}
\affil[4]{Department of Biostatistics, School of Public Health, Brown University, Providence, RI}

\maketitle{}

\thispagestyle{empty}

\clearpage

\textbf{Running title:} Global sensitivity analysis for generalizability and transportability

\clearpage

\setcounter{page}{1}

\vspace*{1in}

\begin{abstract}
\noindent
\linespread{1.3}\selectfont
When individuals participating in a randomized trial differ with respect to the distribution of effect modifiers compared compared with the target population where the trial results will be used, treatment effect estimates from the trial may not directly apply to target population. Methods for extending -- generalizing or transporting -- causal inferences from the trial to the target population rely on conditional exchangeability assumptions between randomized and non-randomized individuals. The validity of these assumptions is often uncertain or controversial and investigators need to examine how violation of the assumptions would impact study conclusions. We describe methods for global sensitivity analysis that directly parameterize violations of the assumptions in terms of potential (counterfactual) outcome distributions. Our approach does not require detailed knowledge about the distribution of specific unmeasured effect modifiers or their relationship with the observed variables. We illustrate the methods using data from a trial nested within a cohort of trial-eligible individuals to compare coronary artery surgery plus medical therapy versus medical therapy alone for stable ischemic heart disease.
\end{abstract}

\clearpage

\section{Background}

When individuals participating in a randomized trial differ with respect to the distribution of effect modifiers compared with the target population where the trial results will be used, treatment effect estimates from the trial may not directly apply to the target population \cite{rothwell2005}. Instead, investigators need appropriate statistical methods to extend -- ``generalize'' or ``transport'' \cite{hernan2016discussionkeiding, dahabreh2019commentaryonweiss} -- causal inferences from the trial to the target population \cite{dahabreh2018generalizing, dahabreh2020transportingStatMed} by combining trial data on baseline covariates, treatments, and outcomes with observational data on baseline covariates from a sample of non-randomized individuals from the target population. The methods allow the identification and estimation of potential (counterfactual) outcome means and treatment effects in the target population or the subset of non-randomized individuals, provided assumptions about the conditional exchangeability of randomized and non-randomized individuals hold. These assumptions are untestable and their validity is often uncertain or controversial. Confidence in the results of analyses extending inferences from the trial to the target population can be better calibrated by conducting sensitivity analyses to examine the impact of violations of the assumptions \cite{robins2000c, robins1999association}. 

Previous proposals for sensitivity analysis methods in the context of analyses extending inferences from a trial to a target population \cite{nguyen2017sensitivity, nguyen2018sensitivity} can be useful when there exists strong background knowledge about unmeasured variables. These methods, however, have several limitations: they rely on assumptions about causal effects at the individual level, sometimes require all effect modifiers to have been measured among trial participants (i.e., unmeasured variables are only allowed in the sample of non-participants), and require the unmeasured variable to be scalar (for all practical purposes) or, even more restictively, binary (for a subset of the methods). In recent work \cite{dahabreh2019sensitivitybiascor}, we described sensitivity analysis methods that do not require such detailed background knowledge or additional covariate data from the randomized individuals. Our proposed methods, however, are most appropriate when the outcome is unbounded, but are less useful for continuous outcomes with bounded support, binary, or count outcomes. 

Here, we propose methods for ``global'' \cite{scharfstein2018globalBiometrics, scharfstein2018globalSMMR} sensitivity analysis that (1) do not require detailed background knowledge about the distribution of specific unmeasured effect modifiers or their relationship with the observed variables; and (2) can be used with all common cross-sectional outcome types (e.g., continuous, binary, or count data). Our approach directly parameterizes violations of the assumptions in terms of the potential outcome distributions among randomized and non-randomized individuals. We illustrate the methods using data from the Coronary Artery Study (CASS), a trial nested within a cohort study of trial-eligible individuals to compare coronary artery surgery plus medical therapy versus medical therapy alone for individuals with stable ischemic heart disease.

\section{Study design and causal estimands}\label{section_contrasts}

We consider a trial embedded, via prospective recruitment or retrospective linkage, in a cohort of individuals who are candidates for the treatments evaluated in the trial; we refer to this design as a nested trial design \cite{dahabreh2021studydesigns}. The cohort is viewed as an independent and identically distributed sample from a superpopulation of substantive interest; we refer to this superpopulation as the \emph{target population} \cite{dahabreh2021studydesigns}. In our motivating application the cohort consists of trial-eligible individuals and thus we focus on generalizability analyses \cite{dahabreh2019commentaryonweiss}. The methods we propose, however, also apply to transportability analyses, where the target population is not the same (usually, broader) than the population of trial-eligible individuals \cite{dahabreh2019commentaryonweiss}. 

Let $X$ denote baseline covariates, $A$ the assigned treatments, $Y$ the outcome measured at the end of the study (binary, continuous, or count), and $S$ the trial participation indicator with $S = 1$ for trial participants and $S = 0$ for non-participants. The data from the nested trial design consist of independent realizations of $(X_i, S_i, S_i A_i, S_i Y_i)$, $i = 1, \ldots, n$, where $n$ is the total number of trial-eligible individuals sampled in the cohort. Note that the data exhibit a special missingness pattern: for randomized individuals we have data on $(S=1, X, A, Y),$ but for non-randomized individuals we only have data on $(S=0, X)$. Throughout, uppercase letters denote random variables, lowercase letters denote realizations of the corresponding random variables, and calligraphic uppercase letters denote the support set of the distribution of the corresponding random variables. 

To define the causal estimands of interest, let $Y^a$ denote the potential (counterfactual) outcome under intervention to set treatment $A$ to $a$, where $a$ belongs in the set of possible treatments $\mathcal A$ \cite{splawaneyman1990, rubin1974,robins2000d}. In the remainder of the paper, we consider binary treatments so that $a \in \mathcal A = \{ 0, 1\}$; extensions to multivalued discrete treatments are straightforward. 

The causal estimand of primary interest is the average treatment effect in the target population of all trial-eligible individuals, $\E [Y^1 - Y^0].$ This estimand is identifiable when the randomized trial is embedded in a sample from the target population (i.e., in nested trial designs) \cite{dahabreh2021studydesigns, dahabreh2018generalizing}.  It is not, in general, equal to the average treatment effect among trial participants, $\E [Y^{1} - Y^{0}] \neq \E [Y^{1} - Y^{0} | S =1] $, when treatment effects vary over baseline covariates that have a different distribution among randomized and non-randomized individuals. Another causal estimand of interest is the average treatment effect in the subset of non-randomized individuals, $\E [Y^1 - Y^{0} | S = 0]$; this estimand is identifiable both in nested trial designs and when the trial and the non-randomized subset of the target population are separately sampled (i.e., in non-nested trial designs) \cite{dahabreh2021studydesigns,dahabreh2020transportingStatMed}.

In this paper paper, we focus on nested trial designs because our substantive example follows that sampling scheme and data structure. Thus, we discuss the identification, estimation, and sensitivity analysis for both $\E[Y^1 - Y^0]$ and $\E [Y^1 - Y^{0} | S = 0]$. Our results for $\E [Y^1 - Y^{0} | S = 0]$, however, can be extended to non-nested trial designs with only minor modifications (see references \cite{dahabreh2021studydesigns,dahabreh2020transportingStatMed} for details).

\section{Identification}

\subsection{Identifiability conditions}\label{section_identifiability}

We now discuss sufficient conditions for identifying the mean of the potential outcomes under each treatment $a$ for the target population of all trial-eligible individuals, $ \E [Y^a]$, and for the subset of trial-eligible non-randomized individuals, $ \E [Y^a  | S =0] $; these potential outcome means are the components of the average treatment effects defined in the previous section. 

\noindent
\emph{I. Consistency of potential outcomes:} 
if $A_i = a$, then $Y_i = Y^{a}_i$, for each $a \in \mathcal A$ and each individual $i$ in the population.

\noindent
\emph{II. Conditional exchangeability over treatment $A$ in the trial:} 
$Y^{a} \mathlarger{\indep} A | (X, S = 1).$ We assume conditional exchangeability of the treatment groups in the trial, so that our results apply to conditionally randomized trials; marginal exchangeability, $Y^{a} \mathlarger{\indep} A| S=1$ or $(Y^{a}, X) \mathlarger{\indep} A| S=1$, is a stronger assumption.

\noindent
\emph{III. Positivity of treatment assignment:} 
$\Pr[A=a | X = x, S=1] > 0$ for each $a \in \mathcal A$ and every $x$ with positive density among randomized individuals, $f_{X|S}(x | S = 1) \neq 0$.

\noindent
\emph{IV. Conditional exchangeability over $S$:} 
$Y^{a} \mathlarger{\indep} S|X$ for each $a \in \mathcal A$; henceforth, we refer to this condition as \emph{conditional generalizability} because in our applied example the target population consists of trial-eligible individuals \cite{dahabreh2019commentaryonweiss}. This condition implies that $\E[Y^a | X, S = 1] = \E[Y^a | X ]$ \cite{dahabreh2018generalizing}, and, for binary $S$, it also implies that $\E[Y^a | X, S = 1] = \E[Y^a | X , S = 0]$ \cite{dahabreh2018generalizing,dahabreh2020transportingStatMed}.

\noindent
\emph{V. Positivity of trial participation:} $\Pr[S=1 | X = x] >0,$ for every $x$ with positive density in the population, $f_X(x) \neq 0$.

Consistency of potential outcomes, conditional exchangeability of the treatment groups in the trial, and positivity of treatment assignment are expected to hold in (marginally or conditionally) randomized trials of well-defined interventions. Our notation implicitly assumes the absence of trial engagement effects (e.g., effects of trial participation on the outcome that are not mediated by treatment assignment) \cite{dahabreh2019identification, dahabreh2020benchmarking}. Furthermore, to simplify exposition, we assume complete adherence to the assigned treatments and no loss to followup. Positivity of trial participation is, in principle, testable, but its evaluation can be challenging in practical applications, particularly when the covariates are high-dimensional \cite{petersen2012diagnosing
}. The conditional generalizability condition, however, is often a strong and untestable assumption; its plausibility needs to be judged on the basis of substantive knowledge when extending inferences from a trial to a target population. Judgement can be aided by the use of causal diagrams, including context-specific single world intervention graphs \cite{dahabreh2019identification} or selection diagrams \cite{bareinboim2016causalfusion}, but the assumption will often remain uncertain, or even controversial, in many applications. 

\subsection{Identification}

When conditions listed in the previous section hold, the potential outcome mean under treatment $a$ in the target population of all trial-eligible individuals, $\E[Y^a]$, can be identified \cite{dahabreh2018generalizing} by the observed data functional
\begin{equation*}
\psi(a) = \E\! \big[\! \E [Y | X, S = 1, A = a] \big] \mbox{, for each } a \in \mathcal A. 
\end{equation*}
Furthermore, the potential outcome mean under treatment $a$ among trial-eligible non-randomized individuals, $\E[Y^a|S=0]$, can be identified \cite{dahabreh2020transportingStatMed} by the observed data functional
\begin{equation*}
\phi(a) = \E\! \big[\! \E [Y | X, S = 1, A = a] | S=0 \big] \mbox{, for each } a \in \mathcal A.
\end{equation*}
Both identification results critically depend on the conditional generalizability condition; we will now discuss global sensitivity analysis methods that can be used to examine the impact of violations of this condition.

\section{Sensitivity analysis for violations of conditional \\ generalizability}\label{sec:sensitivity_analysis}

\subsection{Sensitivity analysis model}

Suppose that the conditional generalizability assumption does not hold, so that $ Y^a \mathlarger{\nindep} S | X, $ that is, for binary $S$, $$ f_{Y^a | X, S}(y | x,  s = 0) \neq f_{Y^a | X,S}(y | x, s = 1).$$

Here, we modify the missing data methods in \cite{scharfstein2018globalSMMR, scharfstein2018globalBiometrics} to parameterize violations of the conditional generalizability assumption using the exponential tilt model, 
\begin{equation}\label{model_exponential_tilt}
	f_{Y^a | X, S}(y | x, s = 0) \propto e^{ \eta_a q(y)} f_{Y^a | X,  S}(y| x , s = 1), \eta_a \in \mathbb R, a \in \mathcal A,
\end{equation} 
where $q$ is a fixed increasing function. Note that $\eta_a = 0$ corresponds to the base case where the conditional generalizability assumption holds; $\eta_a$ values further from zero denote greater potential outcome differences between trial participants and non-participants, conditional on baseline covariates. Because the left-hand-side of equation (\ref{model_exponential_tilt}) is a density, we have that
\begin{equation}\label{model_exponential_tilt_expe}
	\begin{split}
	f_{Y^a | X, S}( y | x, s = 0) 
			&= \dfrac{ e^{\eta_a q(y)} f_{Y^a | X, S}(y| x, s = 1) }{ \E [ e^{\eta_a q(Y^a)} | X, S = 1]}.
	\end{split}
\end{equation}

\subsection{Identification for the sensitivity analysis model}

Under consistency of potential outcomes, exchangeability and positivity of treatment assignment in the trial (i.e., conditions $I$ through $III$), from equation (\ref{model_exponential_tilt_expe}) we obtain
\begin{equation}\label{model_exponential_tilt_expe_factuals}
	\begin{split}
	f_{Y^a | X, S}( y | x, s = 0) = \dfrac{e^{\eta_a q(y)} f_{Y | X, S, A}(y| x, s = 1, a) }{ \E [ e^{\eta_a q(Y)} | X, S = 1 , A = a]}, \eta_a \in \mathbb R, a \in \mathcal A.
	\end{split}
\end{equation}
Because the data are not assumed to contain any outcome information from non-randomized individuals, we cannot nonparametrically identify $f_{Y^a | X, S}(y| x, s = 0)$. Thus, the sensitivity analysis model in equation (\ref{model_exponential_tilt_expe_factuals}) does not have any testable implications for the law of the observed data and $\eta_a$ is not identifiable. In other words, the model is a nonparametric sensitivity analysis model.

In Appendix \ref{appendix:identification}, we show that we can use equation (\ref{model_exponential_tilt_expe_factuals}) to re-express the potential outcome mean under treatment $a$ in the target population, $\E[Y^a]$, as
\begin{equation}\label{eq_mean_identification}
	\psi(a, \eta_a) \equiv \E\! \left[ S \E[Y|X, S=1, A =a] + (1 - S) \dfrac{\E [Y e^{\eta_a q(Y)} | X, S = 1 , A = a]}{\E [e^{\eta_a q(Y)} | X, S = 1 , A = a]} \right].
\end{equation}
Furthermore, we can re-express the potential outcome mean under treatment $a$ in the subset of non-randomized individuals, $\E[Y^a | S = 0]$, as
\begin{equation}\label{eq_mean_identification_S0}
	\phi(a, \eta_a) \equiv \E\! \left[ \dfrac{\E [Y e^{\eta_a q(Y)} | X, S = 1 , A = a]}{\E [e^{\eta_a q(Y)} | X, S = 1 , A = a]} \Big| S = 0 \right].
\end{equation}
As noted, $\eta_a$ is not identifiable from the data; instead, we can use equations (\ref{eq_mean_identification}) and (\ref{eq_mean_identification_S0}) to conduct sensitivity analyses for different, sufficiently dispersed, values of $\eta_a$.

\subsection{Estimation and inference for sensitivity analysis}\label{sec:estimation_inference}

\subsubsection{Outcome model-based sensitivity analysis}

A first approach to sensitivity analysis is to use the sample analogs of equations (\ref{eq_mean_identification}) and (\ref{eq_mean_identification_S0}), for each treatment $a\in\mathcal A$.
For example, when $Y$ is binary and $q$ is the identity function,
\begin{equation} \label{eq:plug_in_binary}
  \begin{split}
\widehat\psi_{\text{\tiny om}}(a, \eta_a) &= \dfrac{1}{n}  \sum\limits_{i=1}^{n}  \left\{  S_i \; \widehat g_a(1| X_i) +  (1 - S_i) \dfrac{ e^{\eta_a } \widehat g_a(1| X_i)}{e^{\eta_a }  \widehat g_a(1| X_i) + \widehat g_a(0| X_i)}  \right\} \mbox{ and} \\[1em]
\widehat\phi_{\text{\tiny om}}(a, \eta_a) &= \left\{ \sum\limits_{i=1}^{n}(1 - S_i) \right\}^{-1} \sum\limits_{i=1}^{n}  \left\{ (1 - S_i) \dfrac{ e^{\eta_a } \widehat g_a(1| X_i)}{e^{\eta_a }  \widehat g_a(1| X_i) + \widehat g_a(0| X_i)}  \right\} ,
  \end{split}
\end{equation}
where $\widehat g_a(y| X)$ is an estimator for $\Pr[Y=y|X, S=1, A =a]$, $y \in \{0,1\}$, and $\eta_a \in \mathbb R$. We refer to the above sensitivity analysis estimators as ``outcome model-based'' because, when $\eta_a = 0$ (i.e., when the generalizability condition holds), they are precisely equivalent to previously described outcome model-based (g-formula \cite{robins1986}) estimators that are valid when the conditional generalizability assumption holds \cite{dahabreh2018generalizing, dahabreh2020transportingStatMed}.

\subsubsection{Estimation for augmented inverse probability or inverse odds weighting sensitivity analysis}

The estimators in the previous section require the estimation of the conditional mean or probability functions for the outcome. Fully nonparametric estimation of these conditional functions is challenging due to the curse of dimensionality, but parametric modeling may be unsatisfactory because of concerns about model misspecification, particularly when $X$ is high-dimensional \cite{robins1997toward}. Flexible modeling, using semiparametric regression or ``machine-learning'' methods, offers a useful compromise. Such flexible modeling can be optimal for estimating the conditional mean or probability functions (e.g., have minimum mean squared prediction error), but flexible modeling of these functions may be associated with substantial finite-sample bias when using sample-analog estimators as in the previous section.

In Appendix \ref{appendix:influence_functions}, we derive the influence functions of $\psi(a, \eta_a)$ and $\phi(a, \eta_a)$ under a non-parameric model for the observed data. We use the influence functions to construct in-sample one-step estimators that should have improved performance compared to the estimators in the previous section. For example, when $Y$ is binary and $q$ is the identity function, for each treatment $a \in \mathcal A$ we construct the following augmented inverse probability weighting sensitivity analysis estimator of $\psi(a, \eta_a)$:
\begin{equation*}
	\begin{split}
\widehat\psi_{\text{\tiny aug}}(a, \eta_a) &= \dfrac{1}{n} \sum\limits_{i=1}^{n} \vast\{ S_i  \left\{ \widehat g_a(1|X_i) + \dfrac{I(A = a )}{\widehat e_a(X_i)} \big\{ Y_i - \widehat g_a(1|X_i) \big\} \right\}  +  \dfrac{(1 - S_i) e^{\eta_a} \widehat g_a(1|X_i)}{e^{\eta_a} \widehat g_a(1|X_i) + \widehat g_a(0|X_i)} \\
&\quad\quad\quad+ \dfrac{I(S_i = 1, A_i = a) \big[1- \widehat p(X_i)\big] e^{\eta_a Y_i} }{\widehat p(X_i) \widehat e_a(X_i) \big[ e^{\eta_a} \widehat g_a(1|X_i) + \widehat g_a(0|X_i) \big]}  \times \Bigg\{ Y_i - \dfrac{e^{\eta_a} \widehat g_a(1|X_i)}{e^{\eta_a} \widehat g_a(1|X_i) + \widehat g_a(0|X_i)}    \Bigg\} \vast\}, 
	\end{split}
\end{equation*}
where, $\widehat g_a(y|X)$ is an estimator for $\Pr[Y = y | X, S = 1 , A = a]$, $y \in \{0,1\}$; $\widehat p(X)$ is an estimator for $\Pr[S=1|X]$; and $\widehat e_a(X)$ is an estimator for $\Pr[A=a|X, S=1]$. Similarly, we construct the following augmented inverse odds weighting estimator of $\phi(a, \eta_a)$:
\begin{equation*}
	\begin{split}
\widehat\phi_{\text{\tiny aug}}(a, \eta_a) &= \left\{ \sum\limits_{i=1}^{n}(1 - S_i) \right\}^{-1}   \sum\limits_{i=1}^{n}  \vast\{  \dfrac{(1 - S_i) e^{\eta_a} \widehat g_a(1|X_i)}{e^{\eta_a} \widehat g_a(1|X_i) + \widehat g_a(0|X_i)} \\
&\quad\quad\quad+ \dfrac{I(S_i = 1, A_i = a) \big[1- \widehat p(X_i)\big] e^{\eta_a Y_i} }{\widehat p(X_i) \widehat e_a(X_i) \big[ e^{\eta_a} \widehat g_a(1|X_i) + \widehat g_a(0|X_i) \big]} \Bigg\{ Y_i  - \dfrac{e^{\eta_a} \widehat g_a(1|X_i)}{e^{\eta_a} \widehat g_a(1|X_i) + \widehat g_a(0|X_i)}  \Bigg\} \vast\}.
	\end{split}
\end{equation*}
We refer to these estimators as ``augmented weighting estimators'' because when $\eta_a = 0$ they are precisely equivalent to perviously described augmented weighting estimators that are valid when when the conditional generalizability assumption holds \cite{dahabreh2018generalizing, dahabreh2020transportingStatMed}.

\subsection{Inference}

When using parametric models, the estimators in the previous section, for fixed $\eta_a$, are (partial) M-estimators and we can use the usual sandwich approach for constructing estimators of their sampling variances \cite{boos2013essential}. In practice, especially when using semiparametric outcome models, it is often convenient to use jackknife \cite{efron1981jackknife} or bootstrap\cite{efron1994introduction} resampling methods to estimate the sampling variance and construct confidence intervals.

\subsection{Relationship with selection models}

Before we apply the sensitivity analysis methods to our motivating example, we consider the relationship between the model in equation (\ref{model_exponential_tilt}) and odds of selection models that are commonly used in missing data problems \cite{rotnitzky1998semiparametric, robins2000c}. 

Using Bayes theorem, under consistency of potential outcomes (condition $I$), and exchangeability in the trial (condition $II$), we can re-write the model in equation (\ref{model_exponential_tilt_expe_factuals}) as
\begin{equation*}
	\begin{split}
	\dfrac{\Pr[S = 0 | X , Y^a = y] }{\Pr[S = 1 | X, Y^a = y] } &=  \dfrac{\Pr[S = 0 | X]}{\Pr[S= 1 | X]}  \times \dfrac{ e^{\eta_a q(y)}}{ \E [ e^{\eta_a q(Y)} | X, S = 1 , A = a]}.
	\end{split}
\end{equation*} 
Taking logarithms,
\begin{equation}\label{model_selection_equiv_to_tilt}
	\begin{split}
	\mbox{logit} \big(\Pr[S = 0 | X , Y^a = y ]\big) &=  \mbox{logit} \big(\Pr[S = 0 | X]\big) + \eta_a q(y) - \ln \E [ e^{\eta_a q(Y)} | X, S = 1 , A = a],
	\end{split}
\end{equation} 
where for a real number $u$, such that $0<u<1$, $\mbox{logit}(u) = \ln \big( u (1 -u)^{-1} \big) .$ In other words, the exponential tilt model for the density of the potential outcomes has an interpretation as an odds of selection model. In Appendix \ref{appendix_selection_models}, we discuss some interesting consequences of the odds of selection parameterization and address the robustness of our estimators to model misspecification.

\section{Sensitivity analysis in CASS} \label{section_example}

\subsection{Data and methods}

\textbf{Study design and data:} CASS was a comprehensive cohort study, comprising a trial nested within a cohort of trial-eligible individuals, that compared coronary artery surgery plus medical therapy (henceforth, ``surgery'') versus medical therapy alone for stable ischemic heart disease \cite{dahabreh2018generalizing}. Details about the CASS are available elsewhere \cite{william1983, investigators1984}. In brief, individuals undergoing angiography in 11 participating institutions were screened for eligibility and the 2099 trial-eligible paindividualstients who met the study criteria were either randomized to surgery or medical therapy (780 randomized individuals), or included in an observational study (1319 non-randomized individuals). Baseline covariates were collected from randomized and non-randomized individuals in an identical manner. No randomized individuals were lost to follow-up in the first 10 years of the study. We did not use information on adherence among randomized individuals, in effect assuming that trial participation and adherence do not have unmeasured common causes.

Of the 2099 trial-eligible individuals, 1688 had complete data on all baseline covariates (733 randomized, 369 in the surgery group and 364 in the medical therapy group; 955 non-randomized). In previous work, to examine whether missing data influenced our results, we performed extensive missing data analyses using multiple imputation and inverse probability of missingness weighting methods. These analyses produced results that were nearly identical to the complete case analyses \cite{dahabreh2018generalizing}. Thus, to focus our exposition on issues related to sensitivity analysis, we only report results based on observations with complete data in this paper.

\textbf{Sensitivity analysis using parametric models:} We used the methods described in Section \ref{sec:sensitivity_analysis} to perform sensitivity analysis in the context of estimating the 10-year risk of death under surgery and medical therapy, and the risk difference and relative risk comparing the treatments, for the population of all trial-eligible individuals and the subset of non-randomized individuals. We used (parametric) logistic regression models for the probability of the outcome, the probability of participation in the trial, and the probability of treatment with the following covariates: age, severity of angina, history of previous myocardial infarction, percent obstruction of the proximal left anterior descending artery, left ventricular wall motion score, number of diseased vessels, and ejection fraction. We chose these variables based on prior analyses of the same data \cite{olschewski1992, dahabreh2018generalizing} and did not perform any model selection.

\textbf{Sensitivity analysis using random forests:} To examine the performance of our sensitivity analysis methods when using more flexible outcome models, we repeated the sensitivity analyses replacing the logistic regression outcome model with a classification random forest procedure. Specifically, we obtained outcome probabilities using a classification random forest with 2000 trees and four input variables (randomly selected for each tree). Our results were similar in stability analyses using random forests of 500 trees (instead of 2000) and using five, six, or seven input variables (results not shown).

\textbf{Choice of sensitivity parameter values:} Under the sensitivity analysis model in equation (\ref{model_exponential_tilt_expe_factuals}), when $Y$ is binary and $q$ is the identity function, for fixed $\eta_a$, the counterfactual conditional outcome probability among non-randomized individuals, $\Pr[Y^a = 1 | X, S = 0]$, is given by the function
\begin{equation*}
  \begin{split}
c(a, \eta_a, X) = \dfrac{e^{\eta_a} g_a(1|X)}{e^{\eta_a}  g_a(1|X) +  g_a(0|X)}.
  \end{split}
\end{equation*}

We set $\eta_1 = - \eta_0 \equiv \eta$ so that when non-randomized individuals have a higher outcome probability under treatment $a=1$ compared to randomized individuals, they also have lower outcome probability under treatment $a=0$. Informally, this represents selection into the trial based on the conditional treatment effect (i.e., selection on unmeasured effect modifiers). With this simplifying assumption, we examined the conditional counterfactual outcome probabilities difference among non-randomized individuals, for $a=1$ and $a=0$, for values of $\eta \in [0,1]$. We used CASS data to estimate the probability of the outcome conditional on the observed covariates among trial participants, $g_a(1|X) = \Pr[Y = 1 | X, S = 1, A = a]$. Appendix Figure \ref{Appendix_Figure_1} shows kernel densities for $\widehat g_1(1|X)$ and $\widehat g_0(1|X)$ among trial participants. Last, we plotted the values of the functions $c(a, \eta, X)$, $a \in \{0,1\}$, at $X$ such that $g_a(1|X)$ was equal to the minimum, 25th percentile, median, mean, 75th percentile, and maximum of the empirical distribution of $\widehat g_a(1|X)$ (Appendix Figures \ref{Appendix_Figure_2} and \ref{Appendix_Figure_3}). Our exploration suggests that the $\eta$ values we chose correspond to fairly strong differential selection into the trial based on the potential outcome, and thus offer a broad enough range of values for sensitivity analysis.

\textbf{Inference:} For all sensitivity analyses -- outcome-model based or augmented weighting; with parametric models or random forests -- we obtained pointwise over $\eta \in [0,1]$ Wald-style 95\% confidence intervals with jackknife standard errors. As a stability analysis, we also obtained percentile intervals using the non-parametric bootstrap with 1000 resamplings.

\subsection{Results}

\begin{figure}[ht!]
	\centering
	\caption{Sensitivity analysis using parametric models for the target population of all trial-eligible individuals in CASS.}	\includegraphics[scale=2]{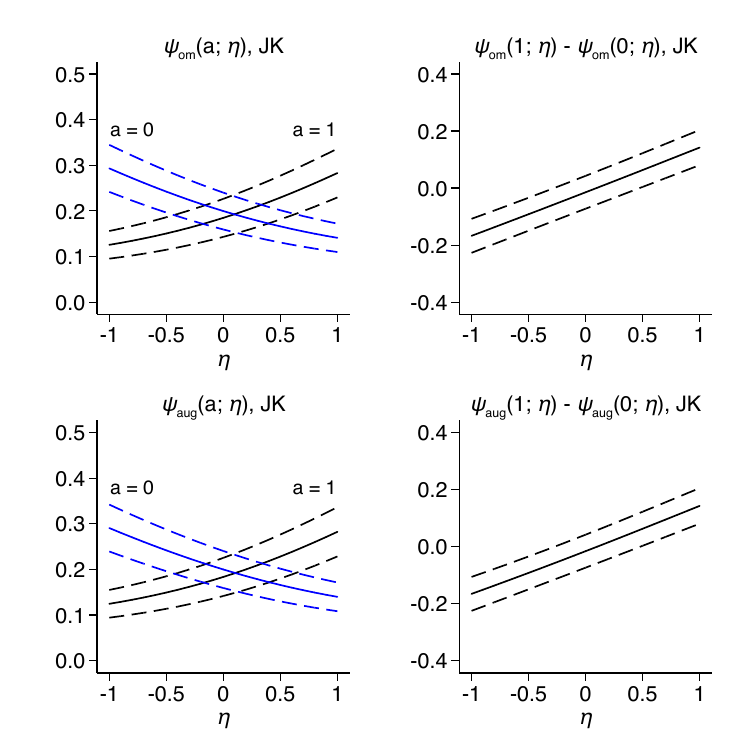}
	\caption*{Solid lines connect point estimates and dashed lines connect pointwise 95\% jackknife (JK) confidence intervals.}
	\label{Fig_1_ALL_parametric_JK}
\end{figure}

\begin{figure}[ht!]
	\centering
	\caption{Sensitivity analysis using parametric models for the subset of trial-eligible non-randomized individuals in CASS.}	\includegraphics[scale=2]{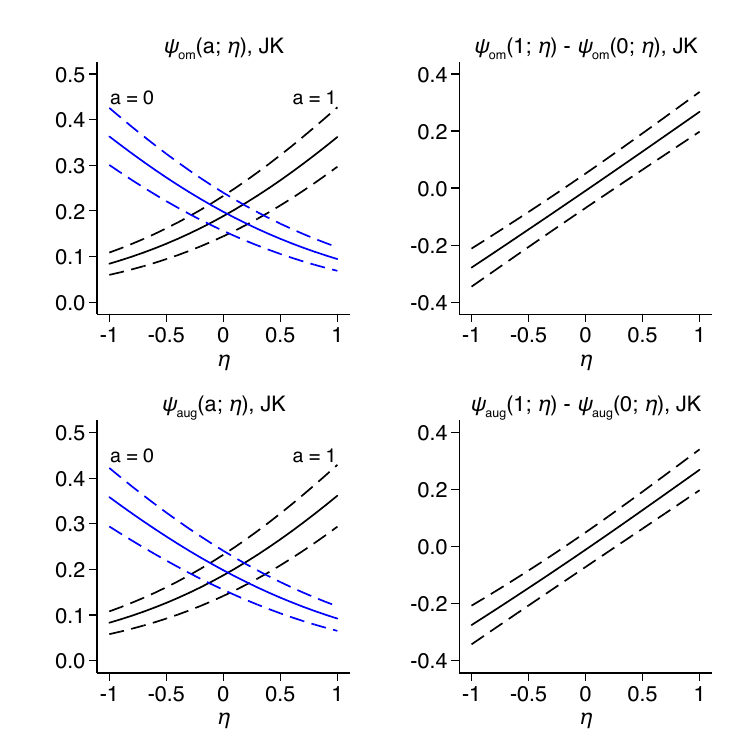}
	\caption*{Solid lines connect point estimates and dashed lines connect pointwise  95\% jackknife (JK) confidence intervals.}
	\label{Fig_2_S0_parametric_JK}
\end{figure}

Figures \ref{Fig_1_ALL_parametric_JK} and \ref{Fig_2_S0_parametric_JK} present sensitivity analysis results for the target population of all trial-eligible individuals and for the subset of non-randomized individuals, using parametric logistic regression models for the probability of trial participation and the conditional probability of the outcome in each treatment group and jackknife confidence intervals. Appendix Figures \ref{Appendix_Figure_4_ALL_RF_JK} and \ref{Appendix_Figure_5_S0_RF_JK} present the same results substituting classification random forests for logistic regression in modeling the conditional probability of the outcome in each treatment group.

Under the base case when the conditional generalizability condition holds (i.e., when $\eta = 0$), outcome-model based and augmented weighting methods produced similar results, both when using parametric models or random forests. Under violations of the conditional generalizability condition, both for the target population of all trial-eligible individuals and the subset of trial-eligible non-randomized individuals, results appeared somewhat less sensitive when using the augmented weighting estimator with random forests compared to using parametric models; differences were most pronounced for extreme $\eta$ values. 

Regardless of the modeling strategy, inferences were more sensitive to violations of the conditional generalizability assumption when considering the potential outcome means and treatment effects in the subset of non-randomized individuals compared to the population of all trial-eligible individuals. This result is expected, because estimates for non-randomized individuals depend entirely on extrapolation (from randomized to non-randomized individuals) whereas estimates for all trial-eligible individuals also include the randomized individuals; inspection of the results in Section \ref{sec:sensitivity_analysis} confirms that terms expressing the ``contribution'' of randomized individuals to the estimators do not depend on the sensitivity parameters $\eta_a$. 

Appendix Figures \ref{Appendix_Figure_6_ALL_parametric_BS} through \ref{Appendix_Figure_9_S0_RF_BS} present results with percentile intervals from the non-parametric bootstrap. Jackknife intervals were slightly wider than bootstrap intervals, particularly when using classification random forests to model the outcome.

\section{Discussion} \label{section_discussion}
A major challenge in generalizability analyses is the need to collect adequate covariate information -- both from trial participants and non-participants -- so that the conditional generalizability condition can be considered to hold. Because the conditional generalizability condition is untestable and background knowledge about effect modification is incomplete and uncertain \cite{dahabreh2016}, investigators need to examine the impact of assumption violations in sensitivity analyses. We described outcome model-based and augmented weighting methods for global sensitivity analysis in studies extending inferences from a trial to the target population of all trial-eligible individuals or the subset of trial-eligible non-randomized individuals. When it is desirable to use flexible outcome models (e.g., machine learning or other semiparametric approaches) to reduce the risk of model misspecification, the outcome model-based sensitivity analysis estimator may exhibit large finite-sample bias because the flexible modeling is tailored towards optimal estimation of the parameters of the outcome model, not the estimation of the sensitivity analysis functionals. In such cases, the augmented weighting sensitivity analysis estimators may perform better.

Our methods are easy to use because they do not require detailed background knowledge about \emph{specific} unmeasured effect modifiers and the relationship between unmeasured effect modifiers and the observed variables. Instead, we directly parameterize violations of the assumptions in terms of the potential outcome distributions of randomized and non-randomized individuals. Our approach has a small number of sensitivity analysis parameters and avoids the difficult task of specifying models for the distribution of the unknown effect modifiers or their relationship with trial participation and the outcome. Intuitively, we exploit the fact that the potential outcomes are not only the ``ultimate confounders'' \cite{greenland1986}, but also the ultimate drivers of effect modification. That is why parameterizing the sensitivity analysis based on the potential outcome distributions leads to fairly simple sensitivity analysis procedures. Alternative methods for sensitivity analysis (e.g., \cite{nguyen2017sensitivity, nguyen2018sensitivity}) may be useful when background knowledge is sharp enough to suggest that a single unmeasured effect modifier with well-understood influence on trial participation and the outcome is responsible for violations of the generalizability condition. In the absence of such sharp knowledge, our methods may be preferred.

As is common with many sensitivity analysis methods that involve counterfactual quantities, it can be challenging to select sensitivity analysis parameters. We described an approach for reasoning about the choice of values for sensitivity analysis parameters in the context of our example. Nevertheless, the preferred ways for eliciting information from substantive experts to parameterize the sensitivity analysis or for sharing such information with end-users remain uncertain. This may be a promising area for future research combining quantitative and qualitative methods.

\clearpage
\bibliographystyle{ieeetr}
\bibliography{Transporting_sensitivity_analysis_GLOBAL}

\clearpage
\section{Figures}

\appendix
\clearpage
\setcounter{page}{1}

\section{Identification of potential outcome means under the sensitivity analysis model}\label{appendix:identification}

Under the conditions of consistency of potential outcomes, and positivity and exchangeability in the trial, the sensitivity analysis model in Section \ref{sec:sensitivity_analysis} can be written as
\begin{equation*}
  f_{Y^a | X, S}(y | x, s = 0) = k(x, a, \eta_a) e^{ \eta_a q(y)} f_{Y | X,  S, A}(y| x , s = 1, a), \eta_a \in \mathbb R, a \in \mathcal A,
\end{equation*} 
where $q$ is an unspecified increasing deterministic function and $k(x, a, \eta_a)$ is the proportionality constant that depends on $x$ and $\eta_a$. For the left-hand-side of the above equation to be a density, 
\begin{equation*}
k(x, a, \eta_a) =  \Big\{ \E \big[ e^{\eta_a q(Y)} | X = x, S = 1, A =a \big] \Big\}^{-1}.
\end{equation*}
These results, and the definition of conditional expectation, imply that,
\begin{equation*}
\E[Y^a | X, S = 0] = \dfrac{\E \big[ Y e^{\eta_a q(Y)} | X, S = 1, A = a \big]}{\E \big[ e^{\eta_a q(Y)} | X, S = 1, A = a \big]}.
\end{equation*}
We now show that $\E[Y^a]$ is identified by the sensitivity analysis functional in equation (\ref{eq_mean_identification}):
\begin{equation*}
  \begin{split}
\E[Y^a] &= \E \big[\E[Y^a | X]\big] \\
&= \E\! \big[\Pr[S=1|X] \E[Y^a | X, S=1]  +  \Pr[S=0|X] \E[Y^a | X, S=0] \big] \\
&= \E\! \big[S \E[Y^a | X, S=1]  + (1 - S) \E[Y^a | X, S=0]  \big] \\
&= \E\! \Bigg[S \E[Y | X, S=1, A =a]  +  (1 - S) \dfrac{\E \big[ Y e^{\eta_a q(Y)} | X, S = 1, A = a \big]}{\E \big[ e^{\eta_a q(Y)} | X, S = 1, A = a \big]}  \Bigg].
  \end{split}
\end{equation*}
Furthermore, $\E[Y^a|S = 0]$ is identified by the sensitivity analysis functional in equation (\ref{eq_mean_identification_S0}): 
\begin{equation*}
  \begin{split}
\E[Y^a | S = 0] &= \E\! \big[\E[Y^a | X, S = 0 ] \big| S = 0\big] \\
&= \E\! \Bigg[ \dfrac{\E \big[ Y e^{\eta_a q(Y)} | X, S = 1, A = a\big]}{\E \big[ e^{\eta_a q(Y)} | X, S = 1, A = a \big]} \Bigg| S = 0  \Bigg].
  \end{split}
\end{equation*}

\clearpage
\section{Influence functions}\label{appendix:influence_functions}

\subsection*{All trial-eligible individuals}

Write the observed data sensitivity analysis functional in equation (\ref{eq_mean_identification}) as 
\begin{equation*}
  \begin{split}
  \psi(a, \eta_a) &= \E_p\! \Bigg[ S \E_p [Y | X, A = a , S =1]  + (1 - S) \dfrac{ \E_p[Y e^{\eta_a q(Y)} | X, S = 1, A = a]  }{ \E_p [e^{\eta_a q(Y)} | X, S = 1, A = a]  }   \Bigg] \\
    &=  \underbrace{ \E_p\!  \big[ S \E_p [Y | X, A = a , S =1]   \big]}_{\psi_1(a)} + \underbrace{\E_p\!  \Bigg[  (1 - S) \dfrac{ \E_p[Y e^{\eta_a q(Y)} | X, S = 1, A = a]  }{ \E_p[e^{\eta_a q(Y)} | X, S = 1, A = a]  }  \Bigg]}_{\psi_2(a, \eta_a)},
  \end{split}
\end{equation*}
where we use the subscript $p$ emphasize dependence on the law of the observed data. 

The influence function for the first term, $\psi_1(a)$, is
\begin{equation*}
\mathit{\Psi}_1^1(a) = S \left\{ \E_{p_0}[Y|X, S = 1 , A = a] + \dfrac{I(A = a )}{\Pr_{p_0}[A = a | X, S = 1] } \Big\{ Y - \E_{p_0}[Y | X, S = 1 , A = a] \Big\} \right\} - \psi_{1p_0}(a),
\end{equation*}
where the zero subscript denotes the ``true'' law. 

Note that $\psi_1(a)$ and $\mathit{\Psi}_1^1(a)$ do not depend on the sensitivity parameters $\eta_a$; they represent the contribution of the randomized individuals.

The influence function for the second term, $\psi_2(a, \eta_a)$, is
\begin{equation*}
	\begin{split}
\mathit{\Psi}_2^1(a, \eta_a) &= (1 - S) \dfrac{\E_{p_0}[Y e^{\eta_a q(Y)} | X, S = 1, A= a]}{\E_{p_0}[e^{\eta_a q(Y)} | X, S = 1, A= a]} + \dfrac{I(S = 1, A = a) \Pr_{p_0}[S = 0 | X]}{\Pr_{p_0}[S=1, A=a | X ] \E_{p_0}[e^{\eta_a q(Y)} | X, S = 1, A= a]} \\
&\quad\quad\quad \times \Bigg\{ Y e^{\eta_a q(Y)}  - \E_{p_0}[Y e^{\eta_a q(Y)} | X, S = 1, A= a]  - \dfrac{\E_{p_0}[Y e^{\eta_a q(Y)} | X, S = 1, A= a]}{\E_{p_0}[e^{\eta_a q(Y)} | X, S = 1, A= a]} \\
&\quad\quad\quad\quad\quad\quad\quad\quad\quad\quad\quad\quad \times \Big\{  e^{\eta_a q(Y)} - \E_{p_0}[e^{\eta_a q(Y)} | X, S = 1, A= a]  \Big\}   \Bigg\} - \psi_{2p_0}(a, \eta_a).
  \end{split}
\end{equation*}

With a bit of algebra, we obtain 
\begin{equation*}
  \begin{split}
\mathit{\Psi}_2^1(a, \eta_a) &= (1 - S) \dfrac{\E_{p_0}[Y e^{\eta_a q(Y)} | X, S = 1, A= a]}{\E_{p_0}[e^{\eta_a q(Y)} | X, S = 1, A= a]} + \dfrac{I(S = 1, A = a) \Pr_{p_0}[S = 0 | X] e^{\eta_a q(Y)}}{\Pr_{p_0}[S=1, A=a | X ] \E_{p_0}[e^{\eta_a q(Y)} | X, S = 1, A= a]} \\
&\quad\quad\quad \times \Bigg\{ Y  - \dfrac{\E_{p_0}[Y e^{\eta_a q(Y)} | X, S = 1, A= a]}{\E_{p_0}[e^{\eta_a q(Y)} | X, S = 1, A= a]}  \Bigg\} - \psi_{2p_0}(a, \eta_a).
	\end{split}
\end{equation*}

It follows that the influence function for $\psi(a, \eta_a)$ is 
$$\mathit{\Psi}^1(a, \eta_a) = \mathit{\Psi}_1^1(a) + \mathit{\Psi}_2^1(a, \eta_a).$$

\noindent
\emph{Remark:} The second term in the expression for $\mathit{\Psi}_2^1(a, \eta_a)$ includes the component $$ \dfrac{\Pr_{p_0}[S = 0 | X] }{\Pr_{p_0}[S=1 | X ] } \times \dfrac{e^{\eta_a q(Y)}}{\E_{p_0}[e^{\eta_a q(Y)} | X, S = 1, A= a]}.$$ It is useful to note that, under our sensitivity analysis model, this component is equal to the odds ratio for trial participation conditional on $X$ and $Y^a$, that is, $$ \dfrac{\Pr_{p_0}[S = 0 | X] }{\Pr_{p_0}[S=1 | X ] } \times \dfrac{e^{\eta_a q(y)}}{\E_{p_0}[e^{\eta_a q(Y)} | X, S = 1, A= a]} = \dfrac{\Pr_{p_0}[S = 0 | X, Y^a = y]}{\Pr_{p_0}[S = 1 | X, Y^a = y]}. $$

\subsection*{Trial-eligible non-randomized individuals}

The influence function of the functional in equation (\ref{eq_mean_identification_S0}) is
\begin{equation*}
	\begin{split}
\mathit{\Phi}^1(a, \eta_a) &= \dfrac{1}{\Pr_{p_0}[S = 0]}  \vast\{ (1 - S) \Bigg\{ \dfrac{\E_{p_0}[Y e^{\eta_a q(Y)} | X, S = 1, A= a]}{\E_{p_0}[e^{\eta_a q(Y)} | X, S = 1, A= a]}  - \phi_{p_0}(a)\Bigg\} \\
&\quad\quad\quad + \dfrac{I(S = 1, A = a) \Pr_{p_0}[S = 0 | X]}{\Pr_{p_0}[S=1, A=a | X ] \E_{p_0}[e^{\eta_a q(Y)} | X, S = 1, A= a]} \\
&\quad\quad\quad\quad\quad\quad \times \Bigg\{ Y e^{\eta_a q(Y)}  - \E_{p_0}[Y e^{\eta_a q(Y)} | X, S = 1, A= a]  - \dfrac{\E_{p_0}[Y e^{\eta_a q(Y)} | X, S = 1, A= a]}{\E_{p_0}[e^{\eta_a q(Y)} | X, S = 1, A= a]} \\
&\quad\quad\quad\quad\quad\quad\quad\quad\quad\quad\quad\quad\quad\quad\quad\quad \times \Big\{  e^{\eta_a q(Y)} - \E_{p_0}[e^{\eta_a q(Y)} | X, S = 1, A= a]  \Big\}   \Bigg\}   \vast\} \\ 
&= \dfrac{1}{\Pr_{p_0}[S = 0]}  \vast\{ (1 - S) \Bigg\{ \dfrac{\E_{p_0}[Y e^{\eta_a q(Y)} | X, S = 1, A= a]}{\E_{p_0}[e^{\eta_a q(Y)} | X, S = 1, A= a]}  - \phi_{p_0}(a)\Bigg\} \\
&\quad\quad\quad + \dfrac{I(S = 1, A = a) \Pr_{p_0}[S = 0 | X] e^{\eta_a q(Y)} }{\Pr_{p_0}[S=1, A=a | X ] \E_{p_0}[e^{\eta_a q(Y)} | X, S = 1, A= a]} \\
&\quad\quad\quad\quad\quad\quad \times \Bigg\{ Y -  \dfrac{\E_{p_0}[Y e^{\eta_a q(Y)} | X, S = 1, A= a]}{\E_{p_0}[e^{\eta_a q(Y)} | X, S = 1, A= a]}   \Bigg\}   \vast\} .
	\end{split}
\end{equation*}

\clearpage
\section{In-sample one-step estimators}\label{appendix:estimators}

\subsection*{All trial-eligible individuals}

Using the results above, for binary $Y$ and $q$ the identity function, we obtain the following one-step in-sample estimator for sensitivity analysis, as a function of $a$ and $\eta_a$:
\begin{equation*}
	\begin{split}
\widehat\psi_{\text{\tiny aug}}(a, \eta_a) &= \dfrac{1}{n} \sum\limits_{i=1}^{n} \vast\{ S_i \left\{ \widehat g_a(1|X_i) + \dfrac{I(A_i = a )}{\widehat e_a(X_i)} \big\{ Y_i - \widehat g_a(1|X_i) \big\} \right\} + \dfrac{(1 - S_i) e^{\eta_a} \widehat g_a(1|X_i)}{e^{\eta_a} \widehat g_a(1|X_i) + \widehat g_a(0|X_i)} \\
&\quad+ \dfrac{I(S_i = 1, A_i = a) \big[1- \widehat p(X_i)\big] e^{\eta_a Y_i} }{\widehat p(X_i) \widehat e_a(X_i) \big[ e^{\eta_a} \widehat g_a(1|X_i) + \widehat g_a(0|X_i) \big]} \times \Bigg\{ Y_i - \dfrac{e^{\eta_a} \widehat g_a(1|X_i)}{e^{\eta_a} \widehat g_a(1|X_i) + \widehat g_a(0|X_i)}  \Bigg\} \vast\}, 
	\end{split}
\end{equation*}
where, $\widehat g_a(y|X)$ is an estimator for $\Pr[Y = y | X, S = 1 , A = a]$,  $a \in \mathcal A$, $y \in \{0,1\}$; $ \widehat p(X)$ is an estimator for $\Pr[S=1|X]$; and $\widehat e_a(X)$ is an estimator for $\Pr[A=a|X, S=1]$, $a \in \mathcal A$.

\subsection*{Trial-eligible non-randomized individuals}

For binary $Y$ and $q$ the identity function, we obtain the following one-step in-sample estimator for sensitivity analysis, as a function of $a$ and $\eta_a$:
\begin{equation*}
	\begin{split}
\widehat\phi_{\text{\tiny aug}}(a, \eta_a) &= \Bigg\{  \sum\limits_{i=1}^{n} (1 - S_i)  \Bigg\}^{-1} \sum\limits_{i=1}^{n} \vast\{  \dfrac{(1 - S_i) e^{\eta_a} \widehat g_a(1|X_i)}{e^{\eta_a} \widehat g_a(1|X_i) + \widehat g_a(0|X_i)} \\
&\quad + \dfrac{I(S_i = 1, A_i = a) \big[1- \widehat p(X_i)\big]  e^{\eta_a Y_i}  }{\widehat p(X_i) \widehat e_a(X_i) \big[ e^{\eta_a} \widehat g_a(1|X_i) + \widehat g_a(0|X_i) \big]  } \times \Bigg\{ Y_i  - \dfrac{e^{\eta_a} \widehat g_a(1|X_i)}{e^{\eta_a} \widehat g_a(1|X_i) + \widehat g_a(0|X_i)} \Bigg\} \vast\},
	\end{split}
\end{equation*}
where, $\widehat g_a(y|X)$, $ \widehat p(X)$, and $\widehat e_a(X)$ are defined as in the previous section.


\clearpage
\renewcommand{\theequation}{D.\arabic{equation}}
\setcounter{equation}{0}

\section{On the relationship between the exponential tilt model and odds of selection models}\label{appendix_selection_models}

\subsection{Re-expression as an odds of selection model}

As noted in the main text, we can re-write the model in equation (\ref{model_exponential_tilt_expe_factuals}) as
\begin{equation*}
  \begin{split}
  \dfrac{\Pr[S = 0 | X , Y^a = y] }{\Pr[S = 1 | X, Y^a = y] } &=  \dfrac{\Pr[S = 0 | X]}{\Pr[S= 1 | X]}  \times \dfrac{ e^{\eta_a q(y)}}{ \E [ e^{\eta_a q(Y)} | X, S = 1 , A = a]}.
  \end{split}
\end{equation*} 
Taking logarithms,
\begin{equation}\label{eq:selection_model_implied}
  \begin{split}
  \mbox{logit} \big(\Pr[S = 0 | X , Y^a = y ]\big) &=  \mbox{logit} \big(\Pr[S = 0 | X]\big) + \eta_a q(y) - \ln \E [ e^{\eta_a q(Y)} | X, S = 1 , A = a],
  \end{split}
\end{equation} 
where for a real number $u$, such that $0<u<1$, $\mbox{logit}(u) = \ln \big( u (1 -u)^{-1} \big) .$ This shows that the exponential tilt model for the density of the potential outcomes has an interpretation as an odds of selection model.

\subsection{Incoherence when using the odds of selection parameterization}\label{appendix_selection_models_incoherence}

We now examine some practical difficulties with using the selection model parameterization directly; our argument follows the logic of \cite{robins2000c} (see pages 39 and 40). In the remainder of this Appendix, we focus on sensitivity analysis for the potential outcome means in the entire target population, $\E[Y^a]$; results for the subset of non-randomized individuals, $\E[Y^a | S = 0]$, are analogous. 

In view of the odds of selection formulation of the sensitivity analysis model, we might think to directly parameterize the sensitivity analysis as
\begin{equation}\label{eq:selection_separate_odds}
\mbox{logit} \Pr[S  = 0 | X , Y^a = y] = m_a(X) + \eta_a q(y), \mbox{ for each } a \in \mathcal A,
\end{equation}
where $m_a(X)$ is such that $m_a(X) = \ln \dfrac{\Pr[S=0 | X]}{\Pr[S = 1 | X] \E[e^{\eta_a q(Y)} | X, S = 1, A= a]}$. In practice, due to the curse of dimensionality, it will often be necessary to posit a model for $m_a(X)$, say $m_a(X) \equiv m_a(X; \gamma_a)$ with finite-dimensional parameter $\gamma_a$ \cite{rotnitzky1998semiparametric, robins2000c}.

To identify the parameters of the selection model above, for fixed $\eta_a$, we can use the population moment conditions 
\begin{equation*}
  \E\! \left[ \dfrac{I(S = 1, A = a)}{ \big\{1 -  \mbox{expit}\big(m_a(X) + \eta_a q(Y) \big) \big\} e_a(X) } \right] = 1 \mbox{ for each } a \in \mathcal A,
\end{equation*}
where $e_a(X) = \Pr[A = a | X, S = 1], a \in \mathcal A $, and for a real number $u$, $\mbox{expit}(u) = \exp (u) \big(1 + \exp(u)\big)^{-1}$. 

Because $e_a(X)$ is under the investigators' control and does not need to be estimated, the above moment conditions can be used to obtain estimates $(\widehat \gamma_1, \widehat  \gamma_0)$, by solving the estimating equations
\begin{equation}\label{eq:estimating_equations_selection_model}
  \sum\limits_{i=1}^{n} \dfrac{\partial m_a(X_i ; \gamma_a) }{\partial\gamma_a}  \left\{ 1 - \dfrac{ I(S_i = 1, A_i = a)}{\big\{1 -  \mbox{expit}\big(m_a(X_i; \gamma_a) + \eta_a q(Y_i) \big) \big\} e_a(X_i)} \right\} = 0 \mbox{ for each } a \in \mathcal A.
\end{equation}
For fixed $\eta_a$ and $q$, we can then estimate $\E[Y^a]$ as 
\begin{equation*}
  \widehat \E [ Y^a] = \dfrac{1}{n} \sum\limits_{i=1}^{n} \dfrac{ I(S_i = 1, A_i = a) Y_i}{\big\{1 -  \mbox{expit}\big( m_a(X_i; \widehat \gamma_a) + \eta_a q(Y_i) \big) \big\} e_a(X_i).}
\end{equation*}

The difficulty is that, in general, there exists no joint distribution for 
\begin{equation*}
(Y^1, Y^0, X, S, A)
\end{equation*}
compatible with our estimates 
\begin{equation*}
(\widehat \E [ Y^1], \widehat \E [ Y^0],  \widehat \gamma_1,  \widehat \gamma_0),
\end{equation*}
because, for fixed $\eta_a$ and $q$, $m_1(X; \gamma_1)$ and $m_0(X; \gamma_0)$ are identified and estimated using different parts of the observed data, leading to potential \emph{incoherence}. More specifically, consider the sample analog of
\begin{equation*}
    \Pr[S = 0 | X = x] =  \int\limits  \Pr[S  = 0 | X =x , Y^{a} = y] f_{Y^a|X}(y|x)dy, \mbox{ for each } a \in \mathcal A,
\end{equation*}
under our odds of selection model; that is,
\begin{equation}\label{eq:constraints}
  \widehat \Pr[S = 0 | X = x] = \int\limits \mbox{expit}\big(m_a(x; \widehat \gamma_a) + \eta_a q(y) \big) \big\} \widehat f_{Y^a|X}(y|x)dy, \mbox{ for each } a \in \mathcal A.
\end{equation}
Here, $\widehat f_{Y^a|X}(y|x)$ is such that 
\begin{equation}\label{eq:constraintMeans}
  \widehat \E [ Y^a] = \int y \widehat f_{Y^a|X}(y|x) dy \widehat f_X(x) dx \mbox{ for each } a \in \mathcal A.
\end{equation}

Because the estimation of $m_1(X; \widehat \gamma_1)$ relies on the $(X, S, S \times A, S \times Y )$ data, but the estimation of $m_0(X; \widehat \gamma_0)$ relies on the $(X, S, S \times (1 - A), S \times Y )$ data, in any finite dataset, it is not possible to guarantee that there exist densities $ \widehat f_{Y^a|X}(y|x)$ such that equation (\ref{eq:constraintMeans}) holds for both $a=1$ and $a=0$, and at the same time
\begin{equation*}
  \int\limits \mbox{expit}\big( m_1(X; \widehat \gamma_1) + \eta_1 q(y)\big) \widehat f_{Y^1|X}(y|X)dy = \int\limits \mbox{expit}\big( m_0(X; \widehat \gamma_0) + \eta_0 q(y)\big) \widehat f_{Y^0|X}(y|X)dy,
\end{equation*}
with probability 1 (these equality constraints are derived from \eqref{eq:constraints}).

As noted in \cite{robins2000c}, some analysts may be willing to overlook the potential incoherence by treating $m_a(X; \gamma_a)$ as an acceptable approximation that can be improved by adaptively using more complex models with increasing sample size. Informally, the incoherence disappears as the sample size increases, provided we use increasingly flexible (data-adaptive) specifications. Other analysts, however, may be troubled by the inability to select models $m_a(X; \gamma_a)$, such that, in a given finite dataset, the estimates are in some sense compatible with the above constraints. 

Regardless of one's attitude towards the potential incoherence with the odds of selection in the model of equation (\ref{eq:selection_separate_odds}), the parameterization of the exponential tilt model in Section \ref{sec:sensitivity_analysis} of the main text avoids this difficulty, while being equivalent to a particular member of the class of non-parametric odds of selection model in equation (\ref{eq:selection_model_implied}).

\subsection{Augmented sensitivity analysis estimator of the potential outcome mean}

If we are willing to ignore the potential for incoherence described in the previous subsection, or if we are exclusively interested in sensitivity analysis for the potential outcome mean under only one particular treatment $a$ in the set $\mathcal A$, we can obtain augmented sensitivity analysis estimators. 

Without loss of generality, suppose that we are exclusively interested in the potential outcome mean under intervention to set treatment to $a=1$. And suppose that we parameterize the sensitivity analysis model using the odds of selection formulation,
\begin{equation}\label{eq:selection_separate_odds_example}
\mbox{logit} \Pr[S  = 0 | X , Y^1 = y] = m_{1}(X) + \eta_1 q(y).
\end{equation}

Using the results in Appendix \ref{appendix:influence_functions}, under the above odds of selection model, the influence function for $\psi(1; \eta_1)$ can be re-written as $$\mathit{\Psi}^1(1, \eta_1) =\mathit{\Psi}_1^1(1) + \mathit{\widetilde{\Psi}}_2^1(1, \eta_1),$$ where the first term on the right-hand-side is, as defined earlier,
\begin{equation*}
\mathit{\Psi}_1^1(1) = S \left\{ \E_{p_0}[Y|X, S = 1 , A = 1] + \dfrac{I(A = 1 )}{\Pr_{p_0}[A = 1 | X, S = 1] } \Big\{ Y - \E_{p_0}[Y | X, S = 1 , A = 1] \Big\} \right\} - \psi_{1p_0}(1),
\end{equation*}
but the second term is replaced by
\begin{equation*}
  \begin{split}
&\mathit{\widetilde{\Psi}}_2^1(1, \eta_1) = (1 - S) b_{p_0}(X) \\
&\quad\quad\quad+ \dfrac{I(S = 1, A = 1)  \Pr_{p_0}[S = 0 | X ] e^{\eta_1 q(Y)}}{ \Pr_{p_0}[S = 1 | X ] \E[e^{\eta_1 q(Y)} | X, S = 1 , A = 1] \Pr_{p_0}[A=1 | X , S = 1 ]  } \big\{ Y  - b_{p_0}(X)  \big\} - \psi_{2p_0}(1, \eta_1),
  \end{split}
\end{equation*}
and we define $$b_{p_0}(X) = \dfrac{\E_{p_0}[Y e^{\eta_1 q(Y)} | X, S = 1, A = 1]}{\E_{p_0}[e^{\eta_1 q(Y)} | X, S = 1, A= 1]}. $$

The above re-expression of the influence function suggests an alternative in-sample one-step estimator of $\psi(1; \eta_1)$,
\begin{equation}\label{eq:doubly_robust_estimator}
  \begin{split}
\widetilde\psi_{\text{\tiny aug}}(1, \eta_1) &= \dfrac{1}{n} \sum\limits_{i=1}^{n} \Bigg\{ S_i \left\{ \widehat g_1(X_i) + \dfrac{I(A_i = 1 )}{ e_1(X_i)} \big\{ Y_i - \widehat g_1(X_i) \big\} \right\} + (1 - S_i) \widehat b(X_i) \\
&\quad+ \dfrac{I(S_i = 1, A_i = 1) e^{ \widehat m_{1}(X_i) + \eta_1 q(Y_i) } }{ e_1(X_i) } \times \big\{ Y_i -  \widehat b(X_i) \big\} \Bigg\}, 
  \end{split}
\end{equation}
where, $\widehat g_1(X) $ is an estimator for $\E[Y | X, S = 1 , A = 1]$; $e_1(X) = \Pr[A=1|X, S=1]$; $ \widehat b(X)$ is an estimator for $\dfrac{\E[Y e^{\eta_1 q(Y)} | X, S = 1, A = 1]}{\E[e^{\eta_1 q(Y)} | X, S = 1, A= 1]}$; and $\widehat m_1(X)$ is an estimator for $\ln \dfrac{\Pr[S=0 | X]}{\Pr[S = 1 | X] \E[e^{\eta_1 q(Y)} | X, S = 1 , A = 1]}$. In practice, $\widehat g_1(X)$ can be obtained using conventional outcome modeling methods; $\widehat b(X)$ can be obtained using a weighted least squares regression of $Y$ on $X$ using weights equal to $e^{\eta_1 q(Y)}$; and $\widehat m_{1}(X)$ can be obtained using the estimating equation strategy in display (\ref{eq:estimating_equations_selection_model}) for $a=1$.

Suppose that $\widehat g_1(X)$, $\widehat b(X)$, and $\widehat m_1(X)$ have well-defined limits, $ g_1^*(X)$, $b^*(X)$, and $m_1^*(X)$, respectively. We will now argue that $\widetilde\psi_{\text{\tiny aug}}(1, \eta_1)$ in equation (\ref{eq:doubly_robust_estimator}) is doubly robust, in the sense that it converges in probability to $\psi(1; \eta_1)$ when \emph{either} $b^*(X) = \dfrac{\E[Y e^{\eta_1 q(Y)} | X, S = 1, A = 1]}{\E[e^{\eta_1 q(Y)} | X, S = 1, A= 1]}$ \emph{or} $m_1^*(X) = \ln \dfrac{\Pr[S=0 | X]}{\Pr[S = 1 | X]\E[e^{\eta_1 q(Y)} | X, S = 1 , A = 1]}$, and regardless of whether $ g_1^*(X) = \E[Y | X, S = 1 , A = 1]$.

First, we note that 
\begin{equation*}
  \begin{split}
\dfrac{1}{n} \sum\limits_{i=1}^{n} S_i \left\{ \widehat g_1(X_i) + \dfrac{I(A_i = 1 )}{ e_1(X_i)} \big\{ Y_i - \widehat g_1(X_i) \big\} \right\} \overset{p}{\longrightarrow} \E \Bigg[  S \left\{ g_1^*(X) + \dfrac{I(A = a)}{e_1(X)} \big\{ Y - g_1^*(X) \big\} \right\}  \Bigg] ,
  \end{split}
\end{equation*}
and by an iterated expectation argument, it is easy to verify that, regardless of whether $g_1^*(X) = \E[Y | X, S = 1 , A = 1]$,
$$ \E \left[ S \left\{ g_1^*(X) + \dfrac{I(A = a)}{e_1(X)} \big\{ Y - g_1^*(X) \big\} \right\} \right] = \E \big[ S \E [Y | X , S = 1 , A = 1]\big],$$
because, by study design, the randomization probability, $e_1(X) = \Pr[A = 1 | X, S = 1]$ is known. Thus, we have that 
\begin{equation}\label{eq:DR_part1}
  \begin{split}
\dfrac{1}{n} \sum\limits_{i=1}^{n} S_i \left\{ \widehat g_1(X_i) + \dfrac{I(A_i = 1 )}{ e_1(X_i)} \big\{ Y_i - \widehat g_1(X_i) \big\} \right\} \overset{p}{\longrightarrow} \E \big[ S \E [Y | X , S = 1 , A = 1]\big].
  \end{split}
\end{equation}

Next, we note that
\begin{equation*}
  \begin{split}
\dfrac{1}{n} \sum\limits_{i=1}^{n} &\Bigg\{ (1 - S_i) \widehat b(X_i) + \dfrac{I(S_i = 1, A_i = 1) e^{\widehat m_{1}(X_i) + \eta_1 q(Y_i)}  }{ e_1(X_i) } \times \big\{ Y_i -  \widehat b(X_i) \big\} \Bigg\} \overset{p}{\longrightarrow} \\
  &\quad\quad\quad \E\left[   (1 - S) b^*(X) + \dfrac{I(S = 1, A = 1) e^{ m_{1}^*(X) + \eta_1 q(Y)}  }{ e_1(X) } \times \big\{ Y -   b^*(X) \big\}  \right], 
  \end{split}
\end{equation*}
and proceed to examine the term on the right-hand-side of the above expression under two possible cases.

\vspace{0.1in}
\noindent
\emph{Case 1:} Suppose that $b^*(X) = \dfrac{\E[Y e^{\eta_1 q(Y)} | X, S = 1, A = 1]}{\E[e^{\eta_1 q(Y)} | X, S = 1, A= 1]}$, but we allow for the possibility that  $m_1^*(X) \neq \ln \dfrac{\Pr[S=0 | X]}{\Pr[S = 1 | X] \E[e^{\eta_1 q(Y)} | X, S = 1, A= 1]}$. 

We have that
\begin{equation*}
  \begin{split}
\E&\left[  \dfrac{I(S = 1, A = 1) e^{ m_{1}^*(X) + \eta_1 q(Y)}  }{ e_1(X) } \times \big\{ Y -   b^*(X) \big\}  \right]  \\
&\quad=\E\left[  \dfrac{I(S = 1, A = 1) e^{ m_{1}^*(X) }  }{ e_1(X) }  Y e^{\eta_1 q(Y)} \right] \\
  &\quad\quad\quad\quad\quad\quad- \E\left[  \dfrac{I(S = 1, A = 1) e^{ m_{1}^*(X) + \eta_1 q(Y)}  }{ e_1(X) } \times \dfrac{\E[Y e^{\eta_1 q(Y)} | X, S = 1, A = 1]}{\E[e^{\eta_1 q(Y)} | X, S = 1, A= 1]}   \right]   \\
&\quad=\E\Big[  e^{ m_{1}^*(X) } \E \big[ Y e^{\eta_1 q(Y)} | X, S = 1 , A = 1 \big] \Pr[S = 1 | X] \Big] \\
  &\quad\quad\quad\quad\quad\quad- \E\left[ e^{ m_{1}^*(X) }  \dfrac{  \E[Y e^{\eta_1 q(Y)} | X, S = 1, A = 1]}{\E[e^{\eta_1 q(Y)} | X, S = 1, A= 1]} \E[e^{\eta_1 q(Y)} | X, S = 1, A= 1]  \Pr[S = 1 | X] \right]  \\
&\quad=0.
  \end{split}
\end{equation*}
Thus, we obtain
\begin{equation*}
  \begin{split}
&\E\Big[   (1 - S) b^*(X) + \dfrac{I(S = 1, A = 1) e^{ m_{1}^*(X) + \eta_1 q(Y)}  }{ e_1(X) } \times \big\{ Y -   b^*(X) \big\}  \Big] = \\
&\quad\quad\quad\quad\quad\quad= \E\big[   (1 - S) b^*(X)   \big] \\
&\quad\quad\quad\quad\quad\quad= \E \left[ (1 - S)  \dfrac{\E[Y e^{\eta_1 q(Y)} | X, S = 1, A = 1]}{\E[e^{\eta_1 q(Y)} | X, S = 1, A= 1]} \right],
  \end{split}
\end{equation*}
and
\begin{equation}\label{eq:DR_part2}
  \begin{split}
\dfrac{1}{n} \sum\limits_{i=1}^{n} &\Bigg\{ (1 - S_i) \widehat b(X_i) + \dfrac{I(S_i = 1, A_i = 1) e^{\widehat m_{1}(X_i) + \eta_1 q(Y_i)}  }{ e_1(X_i) } \times \big\{ Y_i -  \widehat b(X_i) \big\} \Bigg\} \overset{p}{\longrightarrow} \\
  &\quad\quad\quad \E \left[ (1 - S)  \dfrac{\E[Y e^{\eta_1 q(Y)} | X, S = 1, A = 1]}{\E[e^{\eta_1 q(Y)} | X, S = 1, A= 1]} \right], 
  \end{split}
\end{equation}

Combining the results in displays (\ref{eq:DR_part1}) and (\ref{eq:DR_part2}), we have the following result: \\
if $b^*(X) = \dfrac{\E[Y e^{\eta_1 q(Y)} | X, S = 1, A = 1]}{\E[e^{\eta_1 q(Y)} | X, S = 1, A= 1]}$, then 
\begin{equation*}
  \begin{split}
    \widetilde\psi_{\text{\tiny aug}}(1, \eta_1) &\overset{p}{\longrightarrow}  \E \big[ S \E [Y | X , S = 1 , A = 1]\big] +  \E \left[ (1 - S)  \dfrac{\E[Y e^{\eta_1 q(Y)} | X, S = 1, A = 1]}{\E[e^{\eta_1 q(Y)} | X, S = 1, A= 1]} \right] \\
    &= \psi(1; \eta_1).
  \end{split}
\end{equation*}

\vspace{0.1in}
\noindent
\emph{Case 2:} Suppose that $m_1^*(X) = \ln \dfrac{\Pr[S=0 | X]}{\Pr[S = 1 | X] \E[e^{\eta_1 q(Y)} | X, S = 1, A= 1]}$, but we allow for the possibility that $b^*(X) \neq \dfrac{\E[Y e^{\eta_1 q(Y)} | X, S = 1, A = 1]}{\E[e^{\eta_1 q(Y)} | X, S = 1, A= 1]}$. 

We have that 
\begin{equation*}
  \begin{split}
\E&\left[   \dfrac{I(S = 1, A = 1) e^{ m_{1}^*(X) + \eta_1 q(Y)}  }{ e_1(X) } \times \big\{ Y -   b^*(X) \big\}  \right] \\
  &\quad\quad\quad= \E\left[   \dfrac{I(S = 1, A = 1) \Pr[S = 0 | X]  e^{\eta_1 q(Y)} }{  \Pr[S = 1 | X]  e_1(X) \E[e^{\eta_1 q(Y)} | X, S = 1, A= 1]} \times \big\{ Y -   b^*(X) \big\}  \right] \\
  &\quad\quad\quad= \E\left[ \Pr[S = 0 | X] \dfrac{\E[Y e^{\eta_1 q(Y)} | X, S = 1, A = 1]}{\E[e^{\eta_1 q(Y)} | X, S = 1, A= 1]}  \right] - \E \big[ \Pr[S = 0 | X ] b^*(X) \big] \\
  &\quad\quad\quad=  \E\left[ (1 - S) \dfrac{\E[Y e^{\eta_1 q(Y)} | X, S = 1, A = 1]}{\E[e^{\eta_1 q(Y)} | X, S = 1, A= 1]}  \right] -  \E\big[  (1 - S)  b^*(X)  \big].
  \end{split}
\end{equation*}

Thus, we obtain 
\begin{equation*}
  \begin{split}
\E&\left[   (1 - S) b^*(X) + \dfrac{I(S = 1, A = 1) e^{ m_{1}^*(X) + \eta_1 q(Y)}  }{ e_1(X) } \times \big\{ Y -   b^*(X) \big\}  \right] \\
  &\quad\quad\quad\quad\quad\quad\quad\quad\quad\quad\quad\quad\quad\quad\quad\quad\quad\quad= \E\left[ (1 - S) \dfrac{\E[Y e^{\eta_1 q(Y)} | X, S = 1, A = 1]}{\E[e^{\eta_1 q(Y)} | X, S = 1, A= 1]}  \right], 
  \end{split}
\end{equation*}
and, similar to \emph{Case 1}, but under the assumptions about model specification of \emph{Case 2},
\begin{equation}\label{eq:DR_part3}
  \begin{split}
\dfrac{1}{n} \sum\limits_{i=1}^{n} &\Bigg\{ (1 - S_i) \widehat b(X_i) + \dfrac{I(S_i = 1, A_i = 1) e^{\widehat m_{1}(X_i) + \eta_1 q(Y_i)}  }{ e_1(X_i) } \times \big\{ Y_i -  \widehat b(X_i) \big\} \Bigg\} \overset{p}{\longrightarrow} \\
  &\quad\quad\quad \E \left[ (1 - S)  \dfrac{\E[Y e^{\eta_1 q(Y)} | X, S = 1, A = 1]}{\E[e^{\eta_1 q(Y)} | X, S = 1, A= 1]} \right].
  \end{split}
\end{equation}

Combining the results in displays (\ref{eq:DR_part1}) and (\ref{eq:DR_part3}), we have the following result: \\ if $m_1^*(X) = \ln \dfrac{\Pr[S=0 | X]}{\Pr[S = 1 | X] \E[e^{\eta_1 q(Y)} | X, S = 1, A= 1]}$, then 
\begin{equation*}
  \begin{split}
    \widetilde\psi_{\text{\tiny aug}}(1, \eta_1) &\overset{p}{\longrightarrow}  \E \big[ S \E [Y | X , S = 1 , A = 1]\big] +  \E \left[ (1 - S)  \dfrac{\E[Y e^{\eta_1 q(Y)} | X, S = 1, A = 1]}{\E[e^{\eta_1 q(Y)} | X, S = 1, A= 1]} \right] \\
    &= \psi(1; \eta_1).
  \end{split}
\end{equation*}

Taken together, these two cases establish the double robustness of $\widetilde\psi_{\text{\tiny aug}}(1, \eta_1)$; see also \cite{dahabreh2020transportingStatMed} or \cite{robins2008higher} (example 2b; page 352) for related results. 

An informal summary of our results is as follows: using the sensitivity analysis parameterization in equation (\ref{model_exponential_tilt_expe_factuals}) of the main text, we obtain a non-doubly robust estimator that avoids the incoherence described in subsection \ref{appendix_selection_models_incoherence}; alternatively, using the parameterization in equation (\ref{eq:selection_separate_odds_example}), we obtain a doubly robust estimator of the potential outcome mean, but which is subject to potential incoherence, and thus most appropriate if our substantive interest is in the potential outcome mean of only one element of $\mathcal A$. In other words, to avoid the possible incoherence we have to pay the price of not having a doubly robust estimator.


\clearpage

\section{Additional results from CASS analyses}\label{appendix:additional_CASS_results}

\renewcommand\thefigure{\thesection\arabic{figure}}

\setcounter{figure}{0} 

\subsection*{Choice of values for the sensitivity analysis parameter}

\begin{figure}[ht!]
\caption{Predicted outcome probability among randomized individuals, under different interventions. Predictions were obtained from logistic regression models fit separately in each treatment group in the trial; predictions were obtained for all randomized individuals.}
  \centering
    \includegraphics[width=1\textwidth]{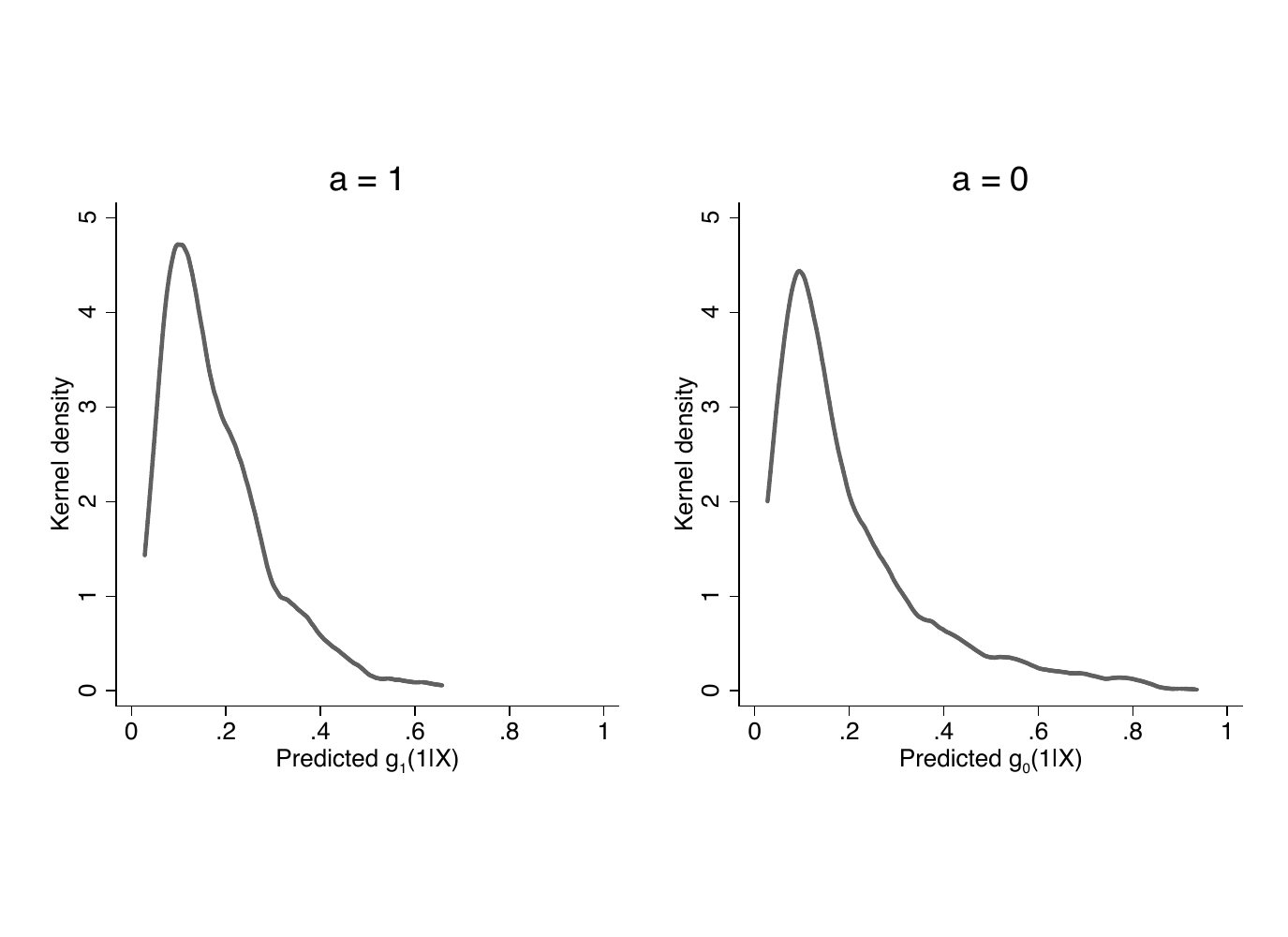}
    \label{Appendix_Figure_1}
\end{figure}

\clearpage
\begin{figure}[ht!]
\caption{Counterfactual probability among $S=0$ under intervention to set $a=1$, for different $\eta$ and $g_1(1|X)$values.}
  \centering
    \includegraphics[width=1\textwidth]{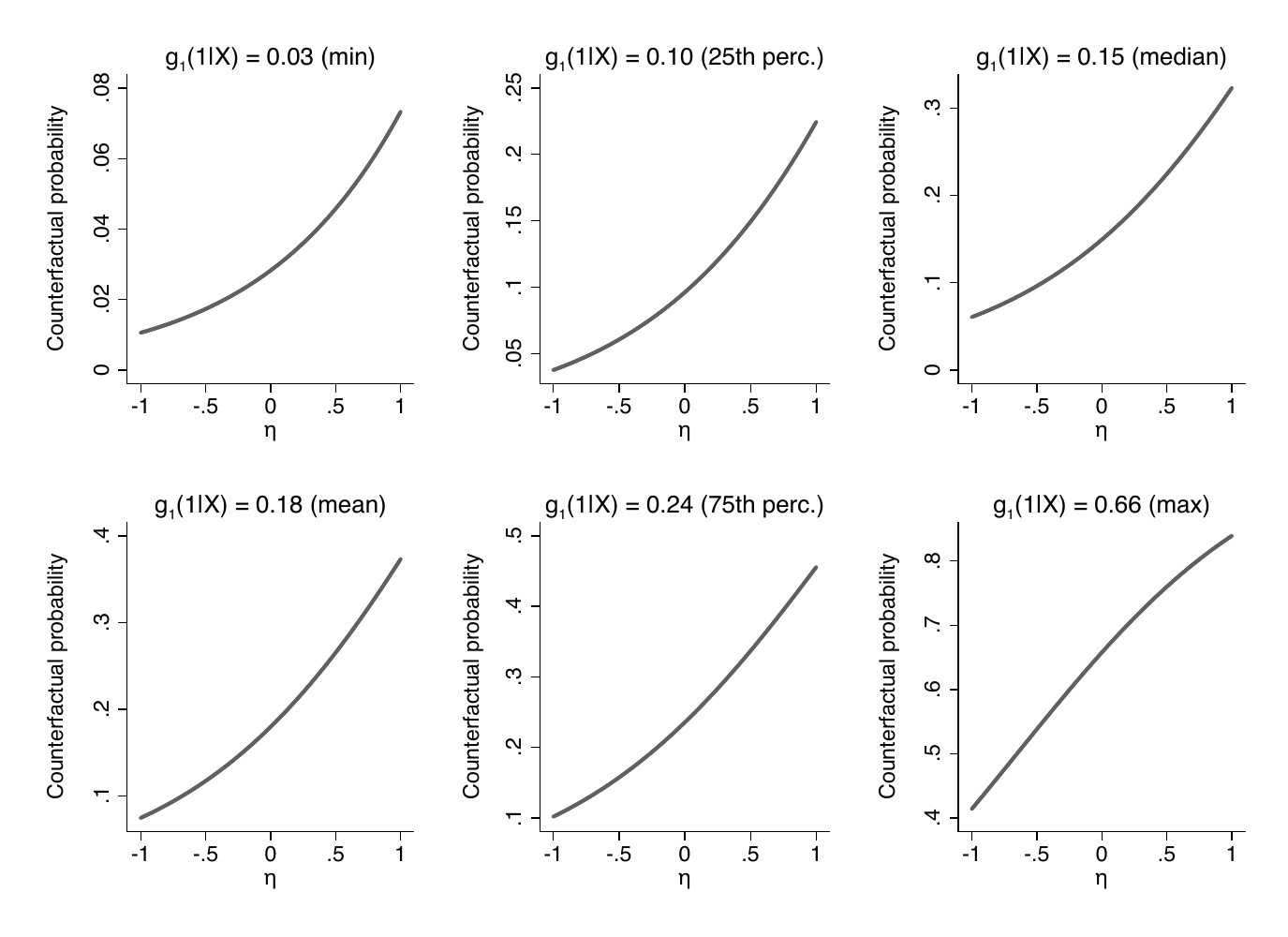}
    \label{Appendix_Figure_2}
\end{figure}

\clearpage
\begin{figure}[ht!]
\caption{Counterfactual probability among $S=0$ under intervention to set $a=1$, for different $\eta$ and $g_0(1|X)$values.}
  \centering
    \includegraphics[width=1\textwidth]{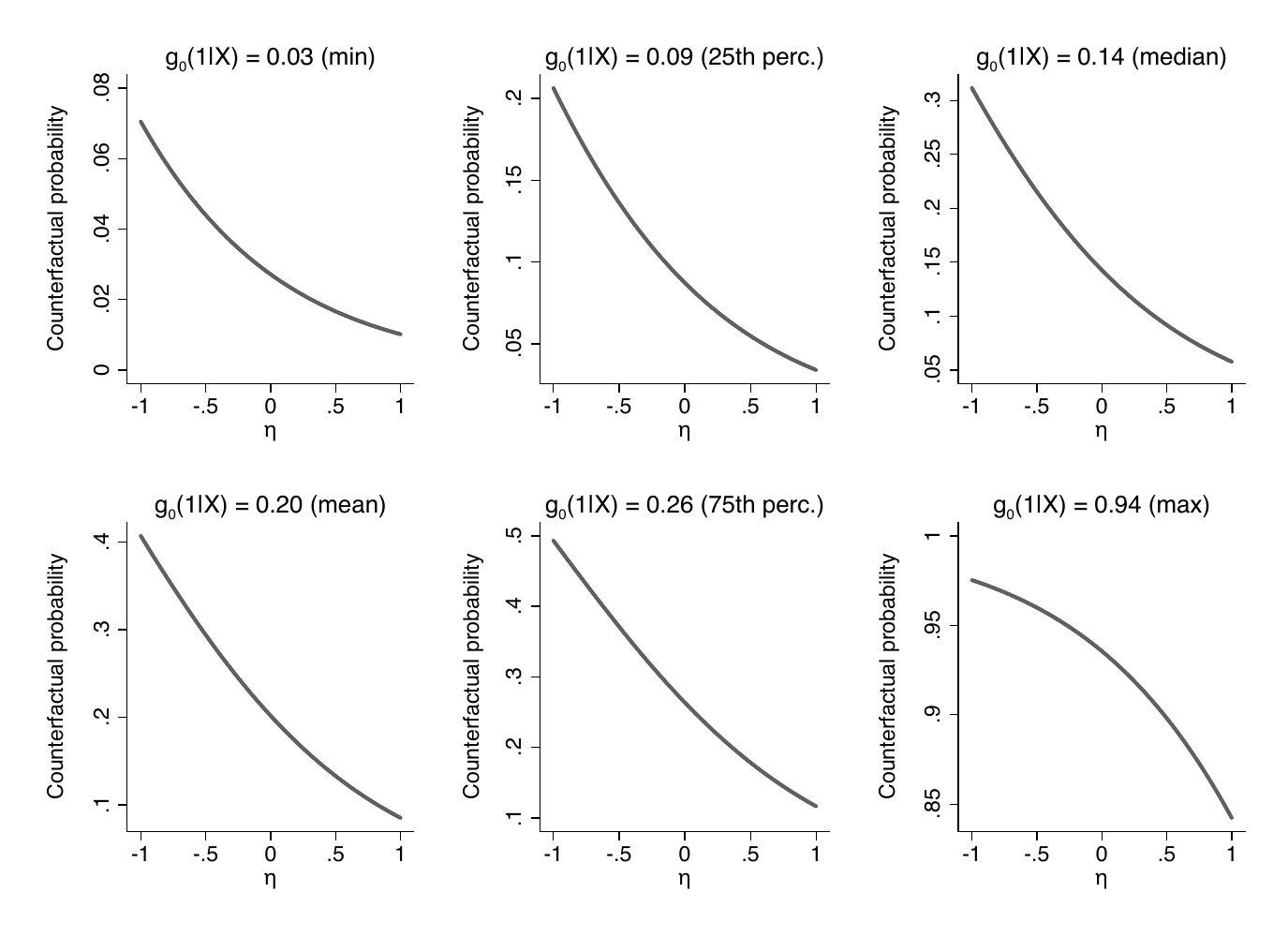}
    \label{Appendix_Figure_3}
\end{figure}

\begin{figure}[ht!]
	\centering
	\caption{Sensitivity analysis using random forest models for the target population of all trial-eligible individuals in CASS.}	\includegraphics[scale=2]{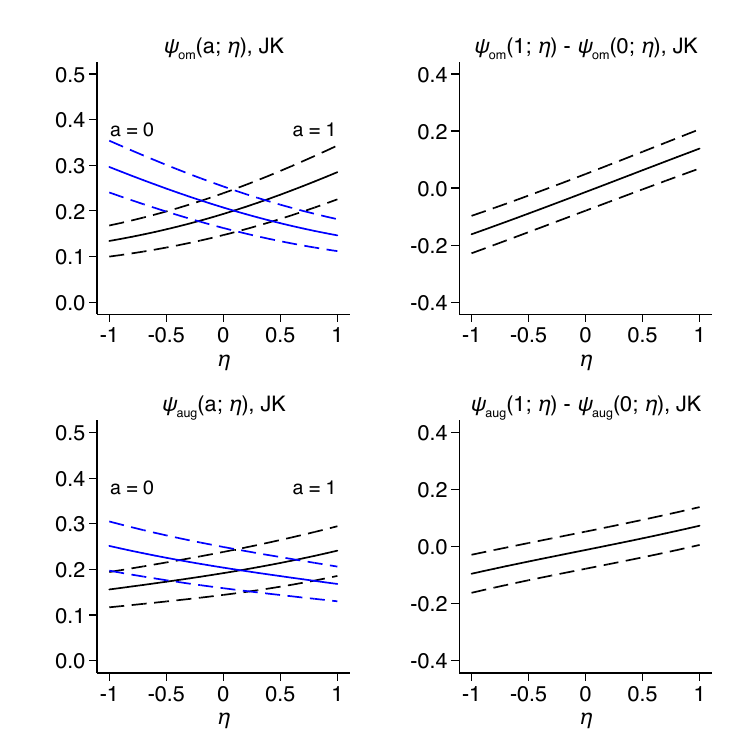}
	\caption*{Solid lines connect point estimates and dashed lines connect pointwise 95\% jackknife (JK) confidence intervals.}
	\label{Appendix_Figure_4_ALL_RF_JK}
\end{figure}

\begin{figure}[ht!]
	\centering
	\caption{Sensitivity analysis using random forest models for the subset of trial-eligible non-randomized individuals in CASS.}	\includegraphics[scale=2]{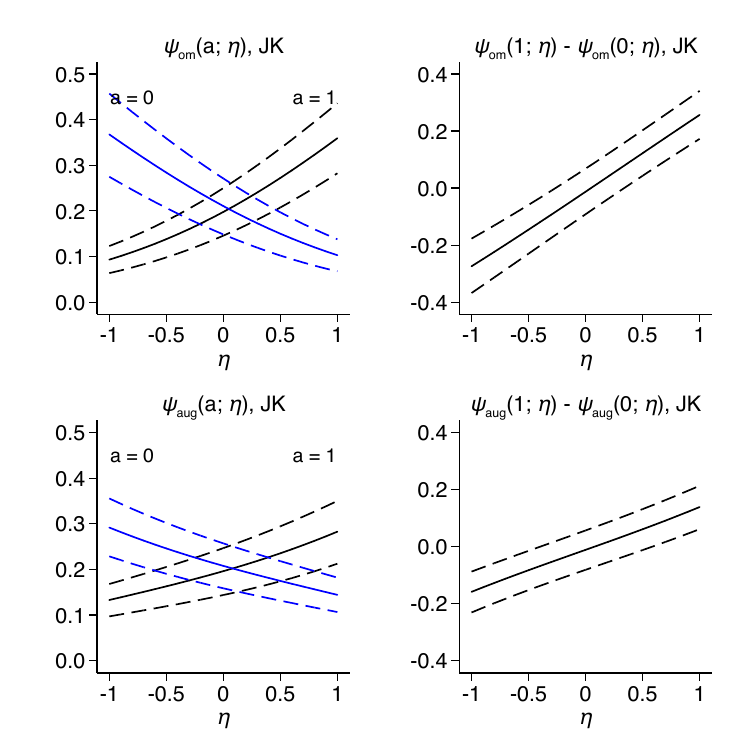}
	\caption*{Solid lines connect point estimates and dashed lines connect pointwise  95\% jackknife (JK) confidence intervals.}
	\label{Appendix_Figure_5_S0_RF_JK}
\end{figure}

\begin{figure}[ht!]
	\centering
	\caption{Sensitivity analysis using parametric models for the target population of all trial-eligible individuals in CASS.}	\includegraphics[scale=2]{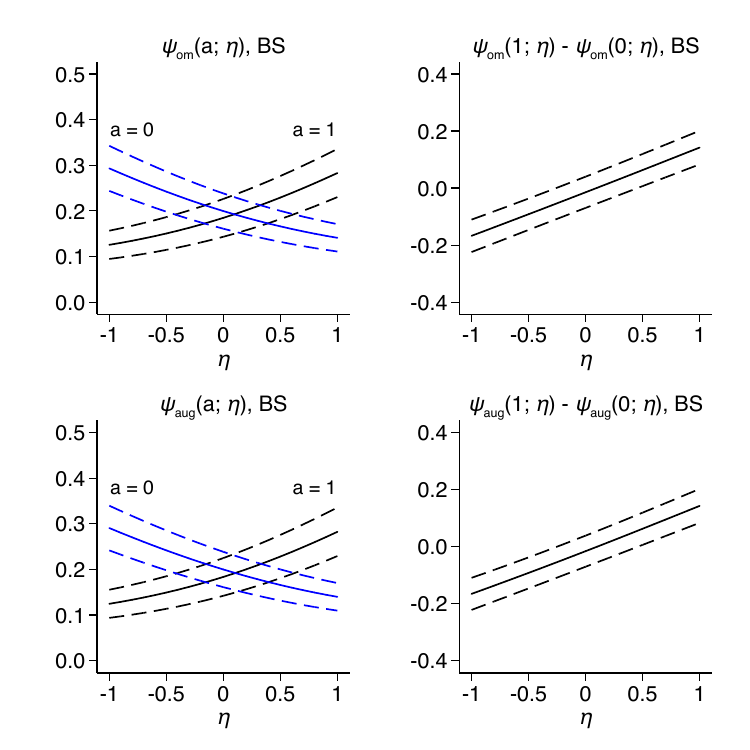}
	\caption*{Solid lines connect point estimates and dashed lines connect pointwise 95\% bootstrap (BS) confidence intervals.}
	\label{Appendix_Figure_6_ALL_parametric_BS}
\end{figure}

\begin{figure}[ht!]
	\centering
	\caption{Sensitivity analysis using parametric models for the subset of trial-eligible non-randomized individuals in CASS.}	\includegraphics[scale=2]{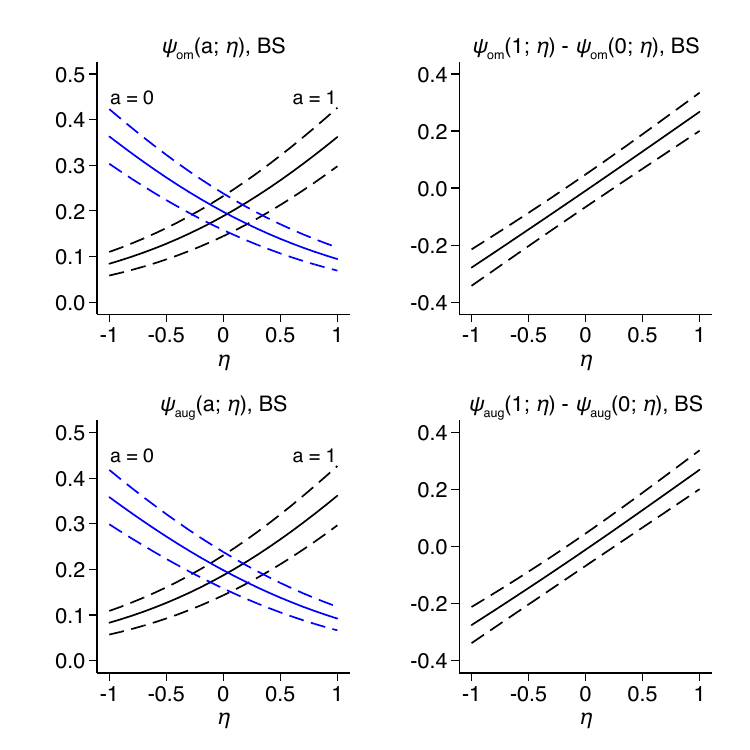}
	\caption*{Solid lines connect point estimates and dashed lines connect pointwise  95\% bootstrap (BS) confidence intervals.}
	\label{Appendix_Figure_7_S0_parametric_BS}
\end{figure}

\begin{figure}[ht!]
	\centering
	\caption{Sensitivity analysis using random forest models for the target population of all trial-eligible individuals in CASS.}	\includegraphics[scale=2]{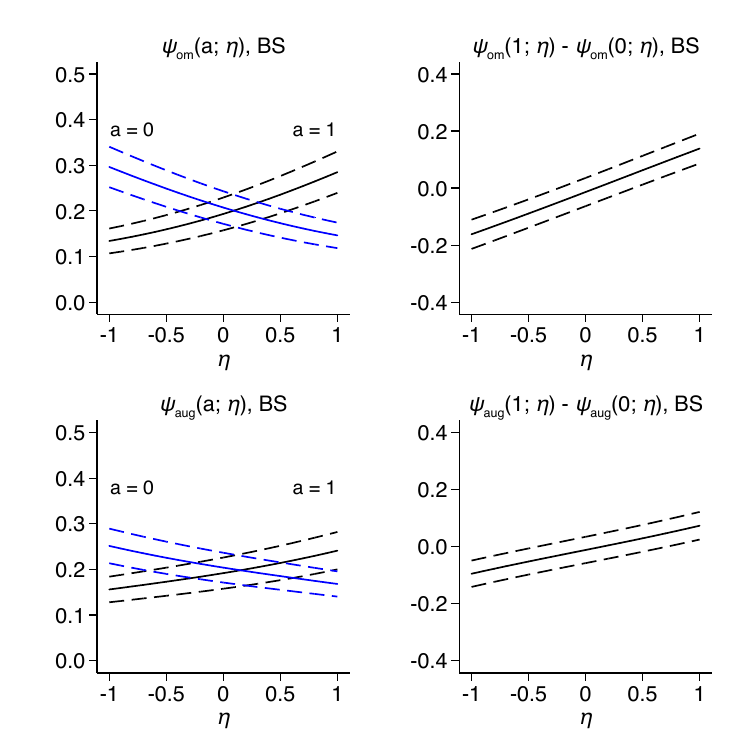}
	\caption*{Solid lines connect point estimates and dashed lines connect pointwise 95\% bootstrap (BS) confidence intervals.}
	\label{Appendix_Figure_8_ALL_RF_BS}
\end{figure}

\begin{figure}[ht!]
	\centering
	\caption{Sensitivity analysis using random forest models for the subset of trial-eligible non-randomized individuals in CASS.}	\includegraphics[scale=2]{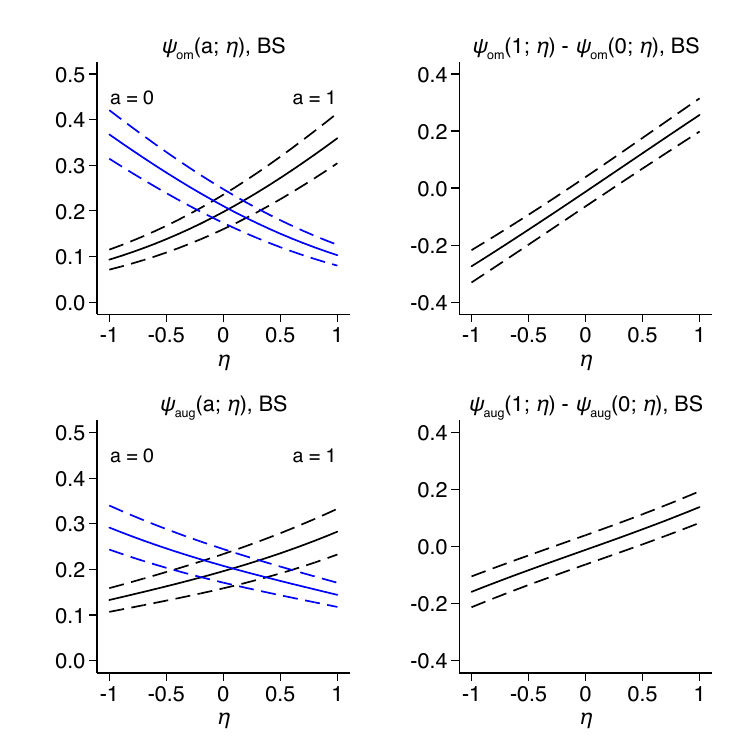}
	\caption*{Solid lines connect point estimates and dashed lines connect pointwise  95\% bootstrap (BS) confidence intervals.}
	\label{Appendix_Figure_9_S0_RF_BS}
\end{figure}


\ddmmyyyydate 
\newtimeformat{24h60m60s}{\twodigit{\THEHOUR}.\twodigit{\THEMINUTE}.32}
\settimeformat{24h60m60s}
\begin{center}
\vspace{\fill}\ \newline
\textcolor{black}{{\tiny $ $generalizability\_failure\_time, $ $ }
{\tiny $ $Date: \today~~ \currenttime $ $ }
{\tiny $ $Revision: \paperversionmajor.\paperversionminor $ $ }}
\end{center}

\end{document}